\documentclass[twocolumn, twocolappendix]{aastex631}
\submitjournal{ApJ}

\shorttitle{Surveys of NH$_{2}$CHO and related species toward massive stars}
\shortauthors{Taniguchi et al.}
\graphicspath{{./}{figures/}}

\begin{document}

\title{Digging into the Interior of Hot Cores with ALMA (DIHCA). III: \\The Chemical Link between NH$_{2}$CHO, HNCO, and H$_{2}$CO}

\correspondingauthor{Kotomi Taniguchi}
\email{kotomi.taniguchi@nao.ac.jp}

\author[0000-0003-4402-6475]{Kotomi Taniguchi}
\affiliation{National Astronomical Observatory of Japan, National Institutes of Natural Sciences, 2-21-1 Osawa, Mitaka, Tokyo 181-8588, Japan}

\author[0000-0002-7125-7685]{Patricio Sanhueza}
\affiliation{National Astronomical Observatory of Japan, National Institutes of Natural Sciences, 2-21-1 Osawa, Mitaka, Tokyo 181-8588, Japan}
\affiliation{Department of Astronomical Science, School of Physical Science, SOKENDAI (The Graduate University for Advanced Studies), Osawa, Mitaka, Tokyo 181-8588, Japan}

\author[0000-0002-8250-6827]{Fernando A. Olguin}
\affiliation{Institute of Astronomy, National Tsing Hua University, Hsinchu 30013, Taiwan}

\author[0000-0003-1602-6849]{Prasanta Gorai}
\affiliation{Department of Space, Earth and Environment, Chalmers University of Technology, SE-412 96, Gothenburg, Sweden}

\author[0000-0003-4615-602X]{Ankan Das}
\affiliation{Institute of Astronomy Space and Earth Science, AJ 316, Salt lake, Sector II, Kolkata 700091, India}

\author[0000-0001-5431-2294]{Fumitaka Nakamura}
\affiliation{National Astronomical Observatory of Japan, National Institutes of Natural Sciences, 2-21-1 Osawa, Mitaka, Tokyo 181-8588, Japan}
\affiliation{Department of Astronomical Science, School of Physical Science, SOKENDAI (The Graduate University for Advanced Studies), Osawa, Mitaka, Tokyo 181-8588, Japan}

\author[0000-0003-0769-8627]{Masao Saito}
\affiliation{National Astronomical Observatory of Japan, National Institutes of Natural Sciences, 2-21-1 Osawa, Mitaka, Tokyo 181-8588, Japan}
\affiliation{Department of Astronomical Science, School of Physical Science, SOKENDAI (The Graduate University for Advanced Studies), Osawa, Mitaka, Tokyo 181-8588, Japan}

\author[0000-0003-2384-6589]{Qizhou Zhang}
\affiliation{Harvard-Smithsonian Center for Astrophysics, 60 Garden St., Cambridge, MA 02138, USA}

\author[0000-0003-2619-9305]{Xing Lu}
\affiliation{Shanghai Astronomical Observatory, Chinese Academy of Sciences, 80 Nandan Road, Shanghai 200030, People's Republic of China}

\author[0000-0003-1275-5251]{Shanghuo Li}
\affiliation{Max Planck Institute for Astronomy, Konigstuhl 17, D-69117 Heidelberg, Germany}

\author[0000-0002-97744-1846]{Huei-Ru Vivien Chen}
\affiliation{Institute of Astronomy, National Tsing Hua University, Hsinchu 30013, Taiwan}

\begin{abstract}
We have analyzed the NH$_{2}$CHO, HNCO, H$_{2}$CO, and CH$_{3}$CN ($^{13}$CH$_{3}$CN) molecular lines at an angular resolution of $\sim 0\farcs3$ obtained by the Atacama Large Millimeter/submillimeter Array (ALMA) Band 6 toward 30 high-mass star-forming regions.
The NH$_{2}$CHO emission has been detected in 23 regions, while the other species have been detected toward 29 regions.
A total of 44 hot molecular cores (HMCs) have been identified using the moment 0 maps of the CH$_{3}$CN line. 
The fractional abundances of the four species have been derived at each HMC.
In order to investigate pure chemical relationships, we have conducted a partial correlation test to exclude the effect of temperature.
Strong positive correlations between NH$_{2}$CHO and HNCO ($\rho=0.89$) and between NH$_{2}$CHO and H$_{2}$CO (0.84) have been found.
These strong correlations indicate their direct chemical links; dual-cyclic hydrogen addition and abstraction reactions between HNCO and NH$_{2}$CHO and gas-phase formation of NH$_{2}$CHO from H$_{2}$CO.
Chemical models including these reactions can reproduce the observed abundances in our target sources.
\end{abstract}


\section{Introduction} \label{sec:intro}

Prebiotic molecules or precursors of organic matter, which are important for life, have been fervently explored in the interstellar medium (ISM) in recent years.
Observations toward the molecular cloud G0.693-0.027 near the Galactic Center have reported the detection of molecules which are related to life; e.g., ethanolamine \citep[NH$_{2}$CH$_{2}$CH$_{2}$OH;][]{2021PNAS..11801314R}, cyanomidyl radical \citep[HCNC;][]{2021MNRAS.506L..79R}, (Z)-1,2-ethenediol \citep[(CHOH)$_{2}$;][]{2022ApJ...929L..11R}.
The detection of these species implies that building blocks of life on Earth may have formed at the early stages of star formation.
However, our current knowledge about the evolution of molecules, from simple molecules through the complex ones to biomolecules, is still lacking.
 
Formamide (NH$_{2}$CHO), the simplest possible amide, has been considered to be a potential prebiotic molecule \citep[e.g.,][]{2019ESC.....3.2122L}. 
This molecule contains a peptide bond that connects amino acids to form proteins.
Its first detection in the ISM was achieved toward the Sagittarius B2 (Sgr B2) high-mass star-forming region \citep{1971ApJ...169L..39R}.
More complex molecules including a peptide bond have been detected in the ISM; e.g., acetamide \citep[CH$_3$CONH$_2$;][]{2006ApJ...643L..25H} and N-methylformamide \citep[CH$_3$NHCHO;][]{2017A&A...601A..49B,2020ApJ...901...37L}.
Formamide could be related to these more complex molecules, thus it is important to reveal their chemistry in the ISM.
Isocyanic acid (HNCO) has been considered as a precursor of NH$_{2}$CHO.
It was firstly discovered in the ISM from Sgr B2 \citep{1972ApJ...177..619S}, soon after the detection of NH$_{2}$CHO.
Although their presence in the ISM has been known for 50 years, their chemistry is still debated. 

Tight correlations between NH$_{2}$CHO and HNCO have been found in observational studies.
\citet{2015MNRAS.449.2438L} derived a power-law relationship of $X$(NH$_{2}$CHO)=0.04$X$(HNCO)$^{0.93}$ from observational results toward young stellar objects (YSOs) with various stellar masses.
\citet{2020ApJ...901...37L} investigated chemical links among several amide molecules toward the high-mass star-forming region NGC\,6334I.
They found that the HNCO/NH$_{2}$CHO abundance ratios in this region are consistent with the average interstellar trend, and suggested that this result strengthens the probable link between HNCO and NH$_{2}$CHO.
\citet{2021ApJ...909..214L} investigated the spatial distribution of several complex organic molecules (COMs) toward the OB cluster-forming region G10.6-0.4, and found a spatial correlation between HNCO and NH$_{2}$CHO.
Such a correlation supports that hydrogenation of HNCO on dust surfaces is a major formation route for NH$_{2}$CHO \citep[e.g.,][]{2014MNRAS.445..151M, 2015MNRAS.449.2438L, 2019ESC.....3.2122L,2016PCCP...1829278S}.

A gas-phase reaction between NH$_{2}$ and H$_{2}$CO has been investigated by quantum calculations and proposed as another formation route of NH$_{2}$CHO \citep{2015MNRAS.453L..31B}.
This reaction has been found to be barrierless and can proceed even in low temperature conditions.
\citet{2017MNRAS.468L...1S} reanalyzed this reaction and proposed an energy barrier of 4.88 K.
As another formation route of NH$_{2}$CHO, solid-phase reactions have been investigated in laboratory experiments \citep[e.g.,][]{2011ApJ...734...78J,2016MNRAS.460.4297F,2019MNRAS.484L.119D,2020ApJ...894...98M} and theoretical studies \citep[e.g.,][]{2018ESC.....2..720R,2019ESC.....3.2158E,2022ApJS..259...39E}.
For example, recent laboratory experiments showed that NH$_{2}$CHO forms in H$_{2}$O ice mixtures containing CO and NH$_{3}$, irradiated by vacuum ultraviolet (VUV) photons \citep{2022ApJ...933..107C}.

Recently, \citet{2022ApJ...937...10L} showed the spatial distribution of NH$_{2}$CHO, HNCO, and H$_{2}$CO, as well as other oxygen-bearing COMs, toward the atmosphere of the HH 212 protostellar disk.
Although they found a similar abundance ratio as the star-forming regions studied in \citet{2019ESC.....3.2122L}, they suggested that HNCO is likely formed in the gas phase and is a daughter molecule of NH$_{2}$CHO.
The spatial distribution of H$_{2}$CO is more extended than that of NH$_{2}$CHO, and the gas phase formation of NH$_{2}$CHO from the reaction between H$_{2}$CO and NH$_{2}$ is questionable from their observational data.

In contrast to the observational studies suggesting the chemical links between NH$_{2}$CHO and HNCO, \citet{2018MNRAS.474.2796Q} suggested that hydrogenation reaction of HNCO is unlikely to produce NH$_{2}$CHO on grain surfaces in their chemical models.
They pointed out that the observed correlations between NH$_{2}$CHO and HNCO may come from the fact that they react to temperature in the same manner rather than a chemical link between them.
\citet{2015A&A...576A..91N} also demonstrated that the hydrogenation reaction of HNCO fails to produce NH$_{2}$CHO efficiently in their laboratory experiments.

\citet{2020ApJ...895...86G} analyzed data obtained with the Atacama Large Millimeter/submillimeter Array (ALMA) Band 4 toward the hot molecular core G10.47+0.03, and investigated the chemistry of molecules containing peptide-like bonds with chemical simulations.
They found that HNCO and NH$_{2}$CHO are linked by the gas-phase dual-cyclic hydrogenation addition and abstraction reactions.
 They also proposed that the main formation route of NH$_{2}$CHO is the reaction between NH$_{2}$ and H$_{2}$CO in the warm-up and post-warm-up phases.
 If the findings of \citet{2020ApJ...895...86G} are applicable to the general picture, we expect to find strong positive correlations between NH$_{2}$CHO and HNCO and between NH$_{2}$CHO and H$_{2}$CO in other star-forming regions.
\citet{2022A&A...668A.109N} investigated the correlation between HNCO and NH$_{2}$CHO using ALMA Band 6 data toward 37 high-mass sources with typical spatial resolutions of 0\farcs5--1\farcs5, corresponding to $\sim1000-5000$ au.
Although they found a correlation between them ($\rho=0.73$), a caveat is that they used column densities which may also depend on the gas mass.
Their observations do not include H$_{2}$CO, and, therefore, the relationship between NH$_{2}$CHO and H$_{2}$CO is still unclear.

In spite of several studies that have investigated the formation processes of NH$_{2}$CHO in star-forming regions, it is still an open question.
In this paper, we present observations the NH$_{2}$CHO, HNCO, H$_{2}$CO, and CH$_{3}$CN ($^{13}$CH$_{3}$CN) lines toward 30 high-mass star-forming regions covered by the `Digging into the Interior of Hot Cores with ALMA (DIHCA)' survey \citep{2021ApJ...909..199O,2022ApJ...929...68O}.
The typical angular resolution is 0\farcs3. 
The source distances are between 1.3 and 5.26 kpc, resulting in linear resolutions of $\sim 520-1590$ au.
Data sets used in this paper are described in Section \ref{sec:data}.
Moment 0 maps of molecular lines and their analyses are presented in Sections \ref{sec:mom0} and \ref{sec:ana}, respectively.
We discuss chemical relationships among NH$_{2}$CHO, HNCO, and H$_{2}$CO by statistical methods in Sections \ref{sec:d1} and \ref{sec:d2}.
The observed molecular abundances are compared with chemical models in Section \ref{sec:d3}.
In Section \ref{sec:conclusion}, we summarize the main conclusions of this paper.

\section{Observations} \label{sec:data}

The 30 high-mass star-forming regions were observed by ALMA in band 6 (226.2 GHz, 1.33 mm) during cycles 4, 5, and 6 between 2016 to 2019 (Project IDs: 2016.1.01036.S, 2017.1.00237.S; PI: Sanhueza).
Observations were performed using the 12\,m array with a configuration similar to C40-5 and more than 40 antennas.
The spectral configuration consists of four spectral windows of 1.8\,GHz of width and a spectral resolution of 976\,kHz ($\sim1.3$\,fkm\,s$^{-1}$).
These windows covered the frequency ranges of 233.5--235.5 GHz, 231.0--233.0 GHz, 216.9--218.7 GHz, and 219.0--221.0 GHz.

The observations were calibrated following the calibration procedure delivered by ALMA with CASA \citep{2022PASP..134k4501C} versions 4.7.0, 4.7.2, 5.1.1-5, 5.4.0-70 and 5.6.1-8.
The data was then self-calibrated in steps of decreasing solution time intervals.
To obtain continuum-subtracted visibilities, we used the procedure from \citet{2021ApJ...909..199O}.
Phase self-calibration was applied to the majority of the continuum-subtracted data sets to produce the data cubes.
The data sets presented in this paper have angular resolutions of 0\farcs3 -- 0\farcs4 with maximum recoverable scales between 8\arcsec -- 11\arcsec.
The source selection criteria are as follows:
\begin{enumerate}
\item The source has a flux density of $>0.1$ Jy at 230 GHz. 
\item The source distance ($d$) is between 1.6 kpc and 3.8 kpc. However, as parallaxes distances have been made available over time, some of the observed targets resulted to be closer, 1.3 kpc, and farther, 5.26 kpc.
\item The clump has an empirical mass-size threshold for high-mass star formation \citep[$M>580$ M$_{\odot}$ ($r$/pc)$^{1.33}$;][]{2010ApJ...723L...7K,2017ApJ...841...97S}.
\end{enumerate}
Further details on the source properties can be found in Ishihara et al. (2023, in prep.).

The imaging of the data cubes from the continuum-subtracted visibilities was performed with the auto-masking routine YCLEAN \citep{2018zndo...1216881C,2018ApJ...861...14C}.
As part of YCLEAN several calls to the CASA task tclean were performed with a multiscale deconvolver and Briggs weighting with a robust parameter of 0.5.
The estimated absolute calibration flux error is 10\%, according to the ALMA Proposer's User Guide\footnote{\url{https://arc.iram.fr/documents/cycle4/ALMA_04_proposer_guide.pdf}}.

\section{Results and Analyses} \label{sec:res}

\subsection{Moment 0 maps} \label{sec:mom0}

\begin{deluxetable*}{llccccc}
\tablecaption{Information on lines used in moment 0 maps \label{tab:molline}}
\tablewidth{0pt}
\tablehead{
\colhead{Species} & \colhead{Transition} & \colhead{Frequency } & \colhead{$E_{\rm {up}}/k$} & \colhead{Binding Energy\tablenotemark{a}} & \colhead{Binding Energy\tablenotemark{b}} & \colhead{Binding Energy\tablenotemark{c}} \\
\colhead{} & \colhead{($J_{K_a, K_c}-J'_{K_a', K_c'}$)} & \colhead{(GHz)} & \colhead{(K)} & \colhead{(K)} & \colhead{(K)} & \colhead{(K)}
}
\startdata
CH$_{3}$CN & $J_{K}=12_{3}-11_{3}$ & 220.709017 & 133.16 & 4680 & 5906 & 4745--7652 \\
NH$_{2}$CHO & $11_{3,9}-10_{3,8}$ & 233.897318 & 94.16 & 5468 & 8104 & 5793--10960 \\
HNCO & $10_{0,10}-9_{0,9}$ & 219.798274 & 58.02 & 4684 & ... & ... \\
H$_{2}$CO & $3_{2, 2}- 2_{2,1}$ & 218.475632 & 68.09 & 2050 & 5187 & 3071--6194 \\
\enddata
\tablenotetext{a}{Values of binding energy were calculated by \citet{2020ApJ...895...86G}.}
\tablenotetext{b}{Values of binding energy were calculated for crystalline ice by \citet{2020ApJ...904...11F}.
\tablenotetext{c}{Values of binding energy were calculated for amorphous nature to mimic the interstellar water ice mantles by \citet{2020ApJ...904...11F}.}}
\end{deluxetable*}

We constructed moment 0 maps of molecular lines. 
At first, we checked the spectra around continuum cores within a single beam size in each high-mass star-forming region and confirmed the detection of each molecular species.
Information on lines used in the moment 0 maps that were checked for detection is summarized in Table \ref{tab:molline}.
These lines are not blended with other lines in most of the target regions.

\begin{figure}[!th]
 \begin{center}
  \includegraphics[bb = 20 20 360 210, scale = 0.72]{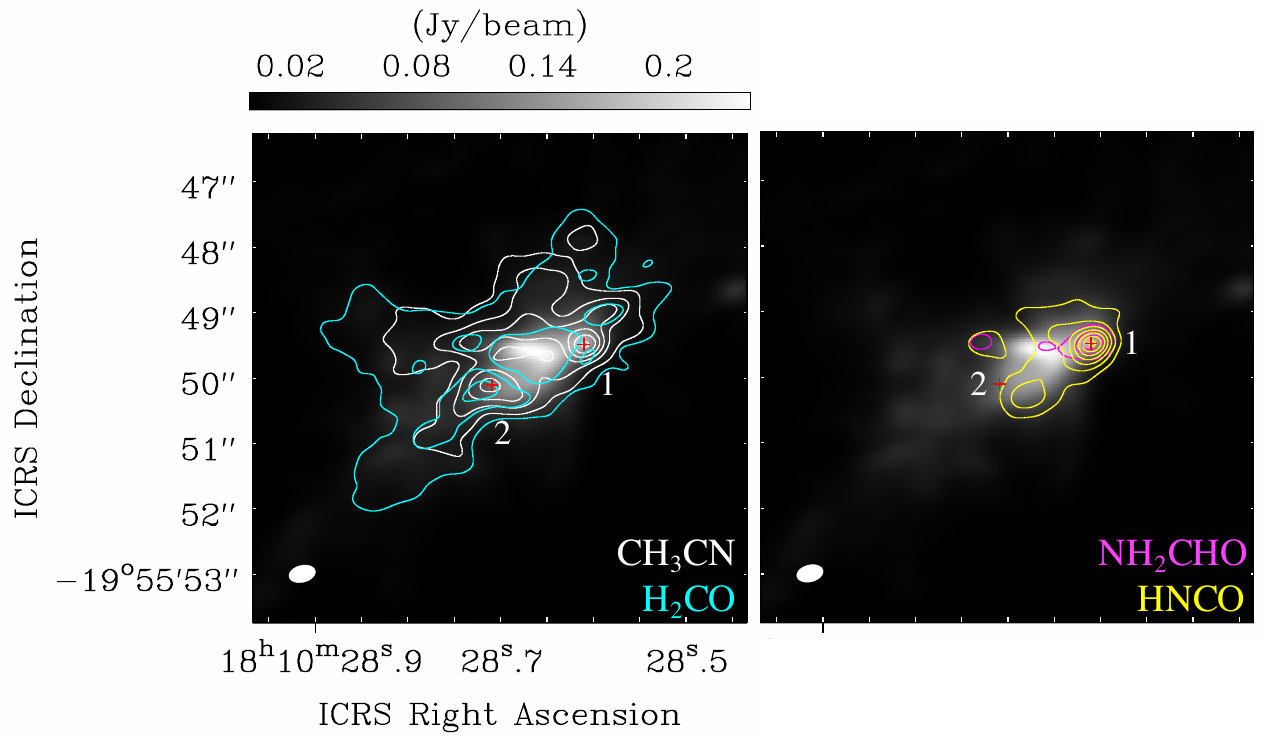}
 \end{center}
\caption{Continuum images (gray scales) overlaid with contours indicating moment 0 maps of molecular lines (left panel: white; CH$_{3}$CN and cyan; H$_{2}$CO, right panel: magenta; NH$_{2}$CHO and yellow; HNCO) toward G10.62-0.38. Red crosses indicate the positions of hot molecular cores (HMCs) identified based on moment 0 maps of the CH$_{3}$CN line. Information on noise levels and contour levels are summarized in Tables \ref{tab:mom0info} and \ref{tab:mom0info2}. \label{fig:momG10}}
\end{figure}

We present moment 0 maps (contour maps) overlaid on the continuum map (gray scale) in Figure \ref{fig:momG10} as an example. 
Moment 0 maps toward all of the regions are presented in Figures \ref{fig:mom01}--\ref{fig:mom04} in Appendix \ref{sec:a1}.
Information on noise levels and contour levels of each map are summarized in Tables \ref{tab:mom0info} and \ref{tab:mom0info2} in Appendix \ref{sec:a1}.

Emission of CH$_{3}$CN has been detected from all of the sources except for IRAS 18337-0743.
In this source, no molecular lines have been detected.
We therefore do not present the data toward this field in this paper.
The detection rate of the CH$_{3}$CN line is 96.7\% ($\frac{29}{30}$).
We identified hot molecular cores (HMCs) based on the moment 0 maps of CH$_{3}$CN. 
If two or more HMCs have been identified in a high-mass star-forming region, we labeled numbers for HMCs in order of integrated intensity, from the highest to the lowest.
We identified 44 HMCs in total over the 29 high-mass star-forming regions.

Emission of the H$_{2}$CO and HNCO lines has been detected from all of the high-mass star-forming regions, except for IRAS 18337-0743.
Emission of NH$_{2}$CHO has been detected from 23 high-mass star-forming regions, and the detection rate is 76.7\% ($\frac{23}{30}$).

Moment 0 maps of all of the molecular lines show almost similar peaks, and they are usually consistent with the continuum peaks within the beam size.
The continuum emission of Band 6 mainly comes from dust emission in most of the sources.
The G5.89-0.37 region is more evolved and shows unique features in both continuum and moment 0 maps.
The continuum emission in this region is dominated by the free-free emission, and shows an explosive dispersal event \citep{2021ApJ...913...29F}.
These results imply that molecules are enhanced in shock regions produced by the expanding motion.

Spatial distributions of NH$_{2}$CHO are the most compact feature compared to the other species. 
This is expected, because the calculated binding energy of NH$_{2}$CHO is the highest among the observed species (see Table \ref{tab:molline}). 
We will discuss this point further in Section \ref{sec:d2}.

\subsection{Analyses} \label{sec:ana}

\begin{deluxetable}{llccc}
\tablecaption{Information on lines used in MCMC analyses \label{tab:analine}}
\tablewidth{0pt}
\tablehead{
\colhead{Species} & \colhead{Transition} & \colhead{Frequency } & \colhead{$E_{\rm {up}}/k$} & \colhead{log$A_{\rm {ij}}$}  \\
\colhead{} & \colhead{$J_{K_a, K_c}-J'_{K_a', K_c'}$} & \colhead{(GHz)} & \colhead{(K)} & \colhead{}
}
\startdata
$^{13}$CH$_{3}$CN & $J_{K}=13_{0}-12_{0}$ & 232.234188 & 78.0 & -2.9667 \\
				   & $J_{K}=13_{1}-12_{1}$ & 232.229822 & 85.2 & -2.9693 \\
				   & $J_{K}=13_{2}-12_{2}$ & 232.216726 & 106.7 & -2.9772 \\
				   & $J_{K}=13_{3}-12_{3}$ & 232.194906 & 142.4 & -2.9907 \\
				   & $J_{K}=13_{4}-12_{4}$ & 232.164369 & 192.5 & -3.0103 \\
				   & $J_{K}=13_{5}-12_{5}$ & 232.125130 & 256.9 & -3.0368 \\
CH$_{3}$CN 		  & $J_{K}=12_{0}-11_{0}$ & 220.747261 & 68.9 & -3.0342 \\
				  & $J_{K}=12_{1}-11_{1}$ & 220.743011 & 76.0 & -3.0372 \\
				  & $J_{K}=12_{2}-11_{2}$ & 220.730261 & 97.4 & -3.0465 \\
				  & $J_{K}=12_{3}-11_{3}$ & 220.709017 & 133.2 & -3.0624 \\
				  & $J_{K}=12_{4}-11_{4}$ & 220.679287 & 183.1 & -3.0857 \\
NH$_{2}$CHO 	  & $11_{3,8}-10_{3,7}$ & 234.316254 & 94.2 & -3.06215  \\	
			         & $11_{3,9}-10_{3,8}$ & 233.897318 & 94.1 & -3.06449 \\
			        & $11_{4,7}-10_{4,6}$ & 233.746504 & 114.9  & -3.09332\\
			        & $11_{4,8}-10_{4,7}$ & 233.735603 & 114.9 & -3.09334 \\
			        & $11_{5,7}-10_{5,6}$ &  233.595412 & 141.7 & -3.13302 \\
HNCO 			& $10_{0,10}-9_{0,9}$	& 219.798320 & 58.0 & -3.82082 \\
				& $10_{1,9}-9_{1,8}$ & 220.585200 & 101.5 & -3.82055 \\
H$_{2}$CO		& $3_{2,2}-2_{2,1}$ & 218.475632 & 68.1 & -3.80403 \\			           			  	   
\enddata
\end{deluxetable}

\begin{figure*}[!th]
 \begin{center}
  \includegraphics[bb = 0 25 479 672, scale=0.7]{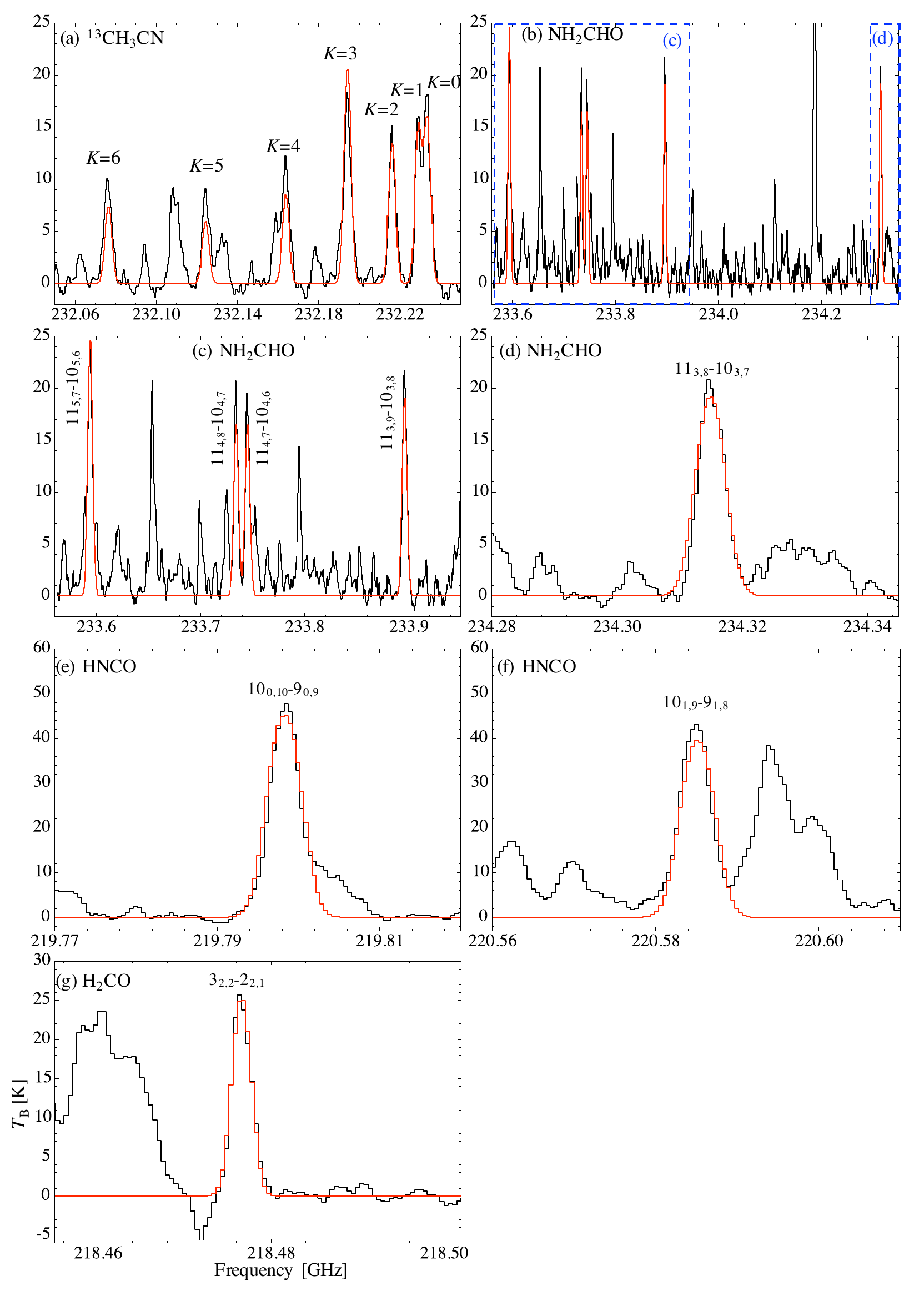}
 \end{center}
\caption{Spectra of molecular lines toward the G10.62-0.38 HMC1; (a) $^{13}$CH$_{3}$CN, (b)--(d) NH$_{2}$CHO (panels (c) and (d) are zoom-in spectra), (e) and (f) HNCO, and (g) H$_{2}$CO. Black and red curves indicate the observed spectra and best-fitting model, respectively. The blue dotted squares in panel (b) indicate the regions for panels (c) and (d), respectively. \label{fig:specG10}}
\end{figure*}

We have conducted line identification and analyzed molecular lines at HMCs with the Markov Chain Monte Carlo (MCMC) method assuming the LTE assumption in the CASSIS software \citep{2015sf2a.conf..313V}.
The positions of HMCs are indicated as red crosses in the moment 0 maps (Figures \ref{fig:mom01}--\ref{fig:mom04}), and their exact positions are summarized in Table \ref{tab:mom0info}.
The upper level energies, Einstein coefficients, degeneracies, partition functions are taken from the CDMS database \citep[for $^{13}$CH$_{3}$CN and CH$_{3}$CN;][]{2016JMoSp.327...95E} and JPL catalogue \citep[for NH$_{2}$CHO, HNCO, H$_{2}$CO;][]{1998JQSRT..60..883P}.
The spectra were obtained within a single beam area. 

In the line identification, we determined representative $V_{\rm {LSR}}$ values at each core using the data of the CH$_{3}$CN lines.
We applied the representative $V_{\rm {LSR}}$ when we identified the lines of the other species, which enables us to pin down molecular lines in line-rich HMCs.

The column density ($N$), excitation temperature ($T_{\rm {ex}}$), line width (FWHM), and systemic velocity ($V_{\rm {LSR}}$) were treated as free parameters in the MCMC method.
The size parameter was fixed, and then the beam filling factor was assumed to be one.
We show the observed spectra (black curves) and the best-fitting model (red curves) toward the G10.62-0.38 HMC1 in Figure \ref{fig:specG10} as an example.
We derived the excitation temperatures ($T_{\rm {ex}}$) and column densities of $^{13}$CH$_{3}$CN by fitting its six $K$-ladder lines ($J_K=13_K-12_K$; $K=0-5$, $E_{\rm {up}}/k=78.0-256.9$ K; see Table \ref{tab:analine}).
We analyzed the $^{13}$CH$_{3}$CN spectra, instead of CH$_{3}$CN, because the low-$K$ lines of the main isotopologue suffer from self absorption in some cases.
The lines of $^{13}$CH$_{3}$CN are found to be optically thin ($\tau < 0.8$).
The $J_{K}=13_{6}-12_{6}$ line of $^{13}$CH$_{3}$CN has been detected in some sources, but this line could not be fitted well with the other lines simultaneously.
This could happen, if emission of this line with a very high upper state energy (335.5 K) comes from smaller regions than the other lines with lower upper state energies.
In general, the best-fitting models underestimate peak intensity of lines with high $E_{\rm {up}}$.
This is likely caused by significant beam dilution effect, where emission of higher $E_{\rm {up}}$ lines arises just in the hotter and inner region and does not fill up the observing beam.
If the lines of $^{13}$CH$_{3}$CN have not been detected (7 cores), we fitted the spectra of the $K$-ladder lines of CH$_{3}$CN ($J_K=12_K-11_K$; $K=0-4$, $E_{\rm {up}}/k=68.9-183.1$ K; Table \ref{tab:analine}).

\begin{figure}[!th]
 \begin{center}
  \includegraphics[bb = 20 30 314 223, scale = 0.84]{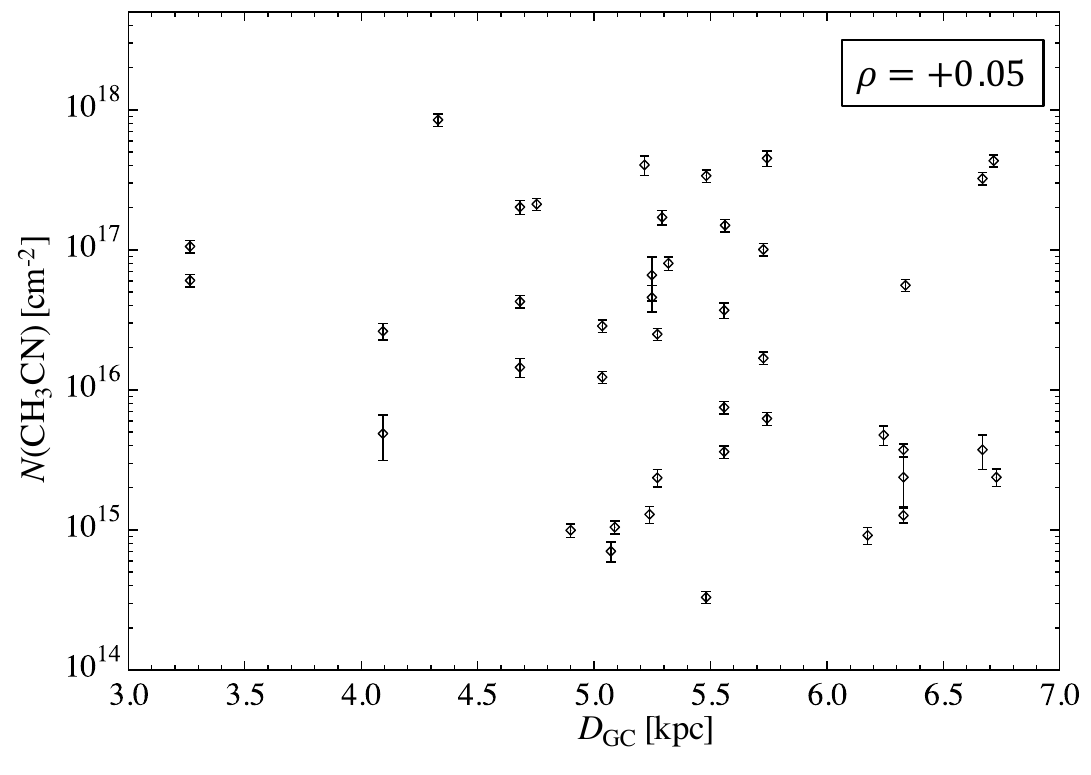}
 \end{center}
\caption{Relationships between the distance from the Galactic Center ($D_{\rm {GC}}$) and the column density of CH$_{3}$CN. \label{fig:Dgc}}
\end{figure}

We converted the column densities of $^{13}$CH$_{3}$CN to those of CH$_{3}$CN using the following formula \citep{2019ApJ...877..154Y}: 
\begin{equation}
^{12}{\rm{C}}/^{13}{\rm{C}} = 5.08 D_{\rm {GC}} + 11.86,
\end{equation}
where $D_{\rm {GC}}$ (in kpc unit) is the distance from the Galactic Center.
The $D_{\rm {GC}}$ values are calculated from the galactic coordinate and the distance between the Sun and the Galactic Center \citep[8 kpc;][]{2003ApJ...597L.121E}.
We summarize the distance between the Sun and the target high-mass star-forming regions ($D$), $D_{\rm {GC}}$, and the calculated $^{12}$C/$^{13}$C ratio in Tables \ref{tab:mcmc} and \ref{tab:cont}. 
Figure \ref{fig:Dgc} shows the relationships between $D_{\rm {GC}}$ and the derived column density of CH$_{3}$CN.
The Spearman's rank correlation coefficient ($\rho$) is derived to be +0.05. 
This means that no artificial effects occurred during the conversion from $N$($^{13}$CH$_{3}$CN) to $N$(CH$_{3}$CN) and the conversion was successful.

Excitation temperature derived from $^{13}$CH$_{3}$CN or CH$_{3}$CN is known to be a good thermometer \citep[e.g.,][]{2014ApJ...786...38H, 2014ApJ...788..187H}.
In the case of the other species, several lines (see Table \ref{tab:analine}) with similar upper-state energies have been clearly detected, and it is difficult to determine their excitation temperatures and column densities simultaneously by fitting their lines.
We then applied the values of excitation temperature derived from the analyses of $^{13}$CH$_{3}$CN or CH$_{3}$CN.
We checked the spectra at all of the positions and selected lines which can be clearly identified and are not affected by contamination by other lines in most positions\footnote{Continuum subtraction could not be done successfully in NGC\,6334I due to line forest. We then did not analyze molecular lines in this field.}.
Table \ref{tab:analine} summarizes information on lines used in the MCMC method.
If the lines of the target species do not show the single Gaussian feature (e.g., possible contamination with other lines), we excluded them from the fitting to derive their column densities precisely.
The derived column densities and excitation temperatures are summarized in Table \ref{tab:mcmc}.
Assuming isothermal gas and using the excitation temperature derived from $^{13}$CH$_{3}$CN or CH$_{3}$CN, we found all other lines to be optically thin.
Results of $V_{\rm {LSR}}$ and FWHM derived by the MCMC method are summarized in Table \ref{tab:vlsr} in Appendix \ref{sec:a2}.

\begin{deluxetable*}{llccccccc}
\tabletypesize{\scriptsize}
\tablecaption{Column densities of each molecule derived by the MCMC method at each position \label{tab:mcmc}}
\tablewidth{0pt}
\tablehead{
\colhead{Region} & \colhead{Position} & \colhead{$T_{\rm {ex}}$} & \colhead{$N$($^{13}$CH$_{3}$CN)} & \colhead{$^{12}$C/$^{13}$C} & \colhead{$N$(CH$_{3}$CN)} & \colhead{$N$(NH$_{2}$CHO)} & \colhead{$N$(HNCO)} & \colhead{$N$(H$_{2}$CO)} \\
\colhead{} & \colhead{} & \colhead{(K)} & \colhead{($\times10^{13}$ cm$^{-2}$)} & \colhead{} & \colhead{($\times10^{14}$ cm$^{-2}$)} & \colhead{($\times10^{14}$ cm$^{-2}$)} & \colhead{($\times10^{14}$ cm$^{-2}$)} & \colhead{($\times10^{15}$ cm$^{-2}$)} 
}
\startdata
G10.62-0.38 & HMC1 & 202.0 (1.2) & 372 (37) & 28 & 1058 (106) & 170 (19) & 719 (72) & 60 (6) \\
& HMC2  & 199.4 (0.4)& 213 (21) & 28 & 606 (61) & ... & 78 (8) & 120 (12) \\
G11.1-0.12 & HMC &  62.1 (0.3) & 7.2 (0.7) & 38 & 11 (1) & 9.7 (1.0) & 23 (2) & 13 (1) \\
G11.92-0.61 & HMC & 212.0 (0.05) & 590 (59) & 36 & 2125 (213) & 198 (22) & 919 (92) & 144 (50) \\
G14.22-0.50 S & HMC & 80.29 (0.03) & ... & 43 & 9.2 (1.3) & 4.7 (0.5) & 19 (2) & 14.9 (1.5) \\
G24.60+0.08 & HMC & 104.9 (0.1) & ...  & 38 & 7.1 (1.2)	& ... & 20 (2) & 15.1 (1.5) \\
G29.96-0.02 & HMC & 270.4 (0.2) & 2510 (251) & 34 & 8489 (849) & 1591 (178) & 5687 (975) & 872 (90) \\
G333.12-0.56 & HMC1 & 193.6 (1.6) & 65 (7) & 39 & 250 (26)& 27 (3) & 110 (11) & 25 (2) \\
& HMC2	& 73.3 (0.3) & 6.11(0.9) & 39 & 24 (3)  & 9.6 (1.0) & 17 (2) & 8.9 (0.9) \\
G333.23-0.06 & HMC1 & 199.99 (0.01) & 81 (11) & 33 & 263 (36) & 68 (7) & 289 (46) & 86 (9) \\
& HMC2 & 100.2 (0.1) & 15 (5) & 33 & 49 (18) & 10.3 (1.6) & 35 (3) & 19 (2) \\
G333.46-0.16 & HMC & 111 (6) & 19 (2) & 40 & 75 (8) & ... &  9.4 (0.9) & 11 (1) \\
& HMC1 & 150 (17) & 93 (11) & 40 & 371 (47) & 42 (4)  & 119 (12) & 16 (6) \\
& HMC2 & 113.3 (0.2) &  9.0 (0.9) & 40 & 36 (4) & 13.0 (1.4) & 52 (5) & 25 (2) \\
G335.579-0.272 & HMC & 249.7 (0.1) & 1055 (164) & 38 & 4047 (629) & 368 (40) & 999 (100) & ... \\
G335.78+0.17 & HMC1 & 249.8 (0.1) & 172 (59) & 39 & 661 (229) & 214 (21) & 766 (90) & 168 (17) \\
& HMC2 & 246 (2) & 119 (16) & 39 & 458 (98) & 86 (9) & 227 (30) & 83 (8) \\
G336.01-0.82 & HMC & 218 (9) & 207 (22) & 39 & 803 (87) & 202 (20) & 619 (62) & ... \\
G34.43+0.24 & HMC & 150.7 (0.3) & 852 (85) & 40 & 3385 (339) & 426	 (47) & 1465 (158) & ... \\		
G34.43+0.24 MM2 &  HMC & 116.6 (0.2) &	& 40 & 3.3 (0.3) & ... &  3.25 (0.3) & 7.6 (0.8) \\
G35.03+0.35A & HMC & 218 (16) & 78 (9) & 44 & 48 (8) & 64 (6) & 238 (66) & 97 (10) \\
G35.13-0.74 & HMC1 & 100 (1) & 54 (8)	& 44 & 37 (4) & 41 (4) & 58 (21) & ... \\	
& HMC2	& 175 (51) &	 ... & 44 & 24 (9) & 10.9 (1.3) & 44 (4) & 24 (2) \\
& HMC3 & 92 (10) & ... & 44 & 12.7 (1.5) & 7.5 (0.7) & 19 (2) & 14 (1) \\
G35.20-0.74 N & HMC & 201.5 (0.7) &127 (13) & 44 & 559 (56) & 145 (15) & 633 (63) & 133 (14) \\
G351.77-0.54  & HMC & 165.3 (0.2) & 943 (94) & 46 & 4336 (434) & 284 (28) & ... & ... \\				
G5.89-0.37 &	HMC1 & 104 (2) & 77 (8) & 37 & 286 (29) 	& ... & 48 (5) & 101 (10) \\
& HMC2 & 99.6 (0.2) & 33 (3) & 37 & 123 (12) & ... & 94 (9) & 82 (8) \\
G343.12-0.06 & HMC & 226 (16) & 441 (52) & 39 & 1710 (201) & 227 (91) & 1440 (196) & 403 (40) \\
IRAS 16562-3959 & HMC1 & 149 (3) & 246 (25) & 41 & 1008 (102) & 54 (5) & 403 (40) & 75 (44) \\
& HMC2 & 88 (8) & 41 (4) & 41 & 169 (17) & 37 (4) & 266 (27) & 26 (3) \\
IRAS 18089-1732 & HMC1 & 247 (21) & 1100 (141) & 41 & 4513 (578) & 470 (48) & 1646 (190) & 201 (20) \\
& HMC2 & 79 (5) & 15 (2) & 41& 62 (7) & 14.0 (1.4) & 39 (4) & 35 (3) \\
IRAS 18151-1208 & HMC & 120 (14) & ... & 38 & 12.9 (1.7) & ... & 17 (2) & 36 (4) \\
IRAS 18182-1433 & HMC1 & 278 (15) & 567 (66) & 36 & 2022 (236) & 391 (93) & 1515 (154) & 327 (30) \\
& HMC2 & 220.1 (0.7) & 120 (12) & 36 & 428 (43) & 40 (8) & 88 (17) & 48 (5) \\
& HMC3	& 138 (20) & 41 (6) & 36 & 145 (23) & ... & 23 (2) & 13.2 (1.3) \\
IRAS 18162-2048	& HMC & 94.3 (0.2) & ... & 46 & 24 (3) & ...	 & 50 (30) & 28 (3) \\
IRDC 18223-1243	 & HMC & 102 (5) & ... & 37 & 10.0 (1.1) & ... & 15 (2)	 & 18 (2) \\					
NGC6334I N & HMC1 & 176 (2) & 707 (71) & 46 & 3235 (324) & 203 (20) & 483 (179) & ... \\		
& HMC2 & 200.7 (0.3) & ... & 46 & 37 (10) & 195 (20) & 343 (122) & 150 (15) \\
W33A & HMC & 199 (5) & 374 (39) & 40 & 1501 (154) & 405 (41) & 3031 (452) & 254 (25) \\
\enddata
\tablecomments{The numbers in parentheses indicate errors including the standard deviation derived from the MCMC analysis and the absolute flux calibration error of 10\%.}
\end{deluxetable*}

We derived the column densities of H$_{2}$, $N$(H$_{2}$), from the continuum data using the following formula \citep[e.g.,][]{2022ApJ...936...80S}:
\begin{equation}
N({\rm {H}}_{2}) = \frac{F_{1.3 {\rm {mm}}} \gamma}{B_{1.3{\rm {mm}} }(T_{\rm {dust}}) \Omega \kappa_{1.3{\rm {mm}} }  \mu_{\rm {H}_2} m_{\rm {H}}}, 
\end{equation}
where $F_{1.3 {\rm {mm}}}$, $\gamma$, $B_{1.3{\rm {mm}}}$($T_{\rm {dust}}$), $\Omega$, $\kappa_{1.3{\rm {mm}}}$, $\mu_{\rm {H}_2}$, and $m_{\rm {H}}$ are continuum peak flux at 1.3 mm, gas-to-dust ratio, the Planck function at 1.3 mm with a dust temperature, the beam solid angle, the H$_{2}$ mean molecular weight (2.8), and the mass of the hydrogen atom, respectively.
We adopted a value of $\kappa_{1.3{\rm {mm}}}=0.9$ cm$^{2}$\,g$^{-1}$ \citep[e.g.,][]{1994A&A...291..943O,2019ApJ...886..102S, 2022ApJ...936...80S}. 
We assumed that the dust temperature is equal to the excitation temperature of $^{13}$CH$_{3}$CN (or CH$_{3}$CN).
The gas-to-dust ratios ($\gamma$) are calculated using the following formula \citep{2017A&A...606L..12G}:
\begin{equation}
{\rm {log}}(\gamma) = 0.087 D_{\rm {GC}} + 1.44.
\end{equation}
The continuum peak flux, gas-to-dust ratios, and derived H$_{2}$ column densities are summarized in Table \ref{tab:cont}.

\begin{deluxetable*}{llccccc}
\tabletypesize{\scriptsize}
\tablecaption{Distances, gas-to-dust ratio, flux density, and H$_{2}$ column density at each position \label{tab:cont}}
\tablewidth{0pt}
\tablehead{
\colhead{Region} & \colhead{Position} & \colhead{$D$} & \colhead{$D_{\rm {GC}}$} & \colhead{$\gamma$} & \colhead{Peak Flux} & \colhead{$N$(H$_{2}$)} \\
\colhead{} & \colhead{} & \colhead{(kpc)} & \colhead{(kpc)} & \colhead{} & \colhead{(mJy beam$^{-1}$)} & \colhead{($\times10^{23}$ cm$^{-2}$)} 
}
\startdata
G10.62-0.38 & HMC1 & 4.95 & 3.26 & 53 & 134 & 22 (2) \\
& HMC2 & 4.95 &	3.26 & 53 & 115	& 20 (2) \\
G11.1-0.12 & HMC & 3.0 & 5.09 & 76 & 7 & 33 (0.3) \\
G11.92-0.61 & HMC & 3.37 & 4.75 & 71 & 71 & 8.6 (0.9) \\
G14.22-0.50 S & HMC & 1.9 & 6.17 & 95 & 25 & 11.3 (1.1) \\ 
G24.60+0.08 & HMC & 3.45 & 5.07 & 76 & 6 & 2.7 (0.3) \\
G29.96-0.02 & HMC & 5.26 & 4.33 & 66 & 96 & 15.0 (1.5) \\
G333.12-0.56 & HMC1 & 3.3 & 5.27 & 79 & 27 & 4.0 (0.4) \\
& HMC2& 3.3 & 5.27 & 79 & 14 & 5.9 (0.6) \\
G333.23-0.06	 & HMC1 & 5.2 & 4.09	& 63 & 30 & 3.4 (0.3)  \\
& HMC2 & 5.2 & 4.09	& 63 & 80 & 19 (2) \\
G333.46-0.16 & HMC & 2.9 & 5.56 & 84 & 45 & 12.8 (1.3) \\
& HMC1 & 2.9 & 5.56 & 84 & 42 & 8.8 (0.9) \\
& HMC2 & 2.9 & 5.56 & 84 & 7 & 1.7 (0.2) \\
G335.579-0.272 & HMC & 3.25 & 5.22 & 78 & 274 & 31 (3) \\
G335.78+0.17 & HMC1 & 3.2 & 5.25 & 79 & 100 & 11.4 (1.1) \\
& HMC2	 & 3.2 & 5.25 & 79 & 30 & 3.5 (0.4) \\
G336.01-0.82 & HMC & 3.1 & 5.32 & 80 & 40	 & 5.4 (0.5) \\
G34.43+0.24 & HMC & 3.5 &5.48 & 83 & 236 & 48 (5) \\
G34.43+0.24 MM2 & HMC & 3.5 & 5.48 & 83 & 16 & 4.2 (0.4) \\
G35.03+0.35A & HMC &	2.32 & 6.24 & 96 & 33 & 5.2 (0.5) \\ 
G35.13-0.74	& HMC1	& 2.2 & 6.33	& 98 & 40 & 14.6 (1.5) \\
& HMC2	& 2.2 & 6.33	& 98 & 29 & 5.9 (0.6) \\
& HMC3 & 2.2 & 6.33	& 98 & 8 & 3.0 (0.3) \\
G35.20-0.74 N & HMC & 2.19 & 6.34 & 98 & 77 & 13.7 (1.4) \\
G351.77-0.54 & HMC & 1.3 & 6.72 & 106 & 158 & 37 (4)  \\
G5.89-0.37 & HMC1 & 2.99 & 5.04 & 76 & 54 & 14.7 (1.5) \\
& HMC2 & 2.99 & 5.04 & 76 & 14 & 3.9 (0.4) \\
G343.12-0.06 & HMC & 2.9 & 5.29 & 80 & 168 & 22 (0.2) \\
IRAS 16562-3959	& HMC1& 2.38 & 5.73 & 87 & 10 &  4.0 (0.4) \\
& HMC2 & 2.38 &	5.73 & 87 & 155 & 102 (10) \\	
IRAS 18089-1732 & HMC1 & 2.34 & 5.74 & 87 & 157 & 20 (2) \\ 
& HMC2 & 2.34 & 5.74 & 87 & 26 & 11 (1) \\
IRAS 18151-1208	 & HMC & 3	.0 & 5.24 & 79 & 8 & 3.5 (0.4) \\
IRAS 18182-1433 & HMC1 &3.58 & 4.68 & 70 & 21 & 3.4 (0.3) \\
& HMC2 & 3.58 & 4.68 & 70 & 23	& 4.8 (0.5) \\
& HMC3 & 3.58 & 4.68 & 70 & 32	& 11 (1) \\ 
IRAS 18162-2048 & HMC & 1.3 & 6.73 & 106 & 319 & 134 (13) \\
IRDC 18223-1243	 & HMC & 3.4 &4.90 & 73 & 7 & 3.3 (0.3) \\
NGC6334I & HMC1 &1.35 & 6.67 & 105 & 266 & 121 (12) \\ 
&HMC2 & 1.35 & 6.67 & 105 & 635 & 290 (29) \\	
NGC6334I N & HMC1 & 1.35 & 6.67 & 105 & 141 & 55 (5) \\
& HMC2	& 1.35 & 6.67 & 105 & 34 & 11 (1) \\ 
W33A & HMC & 2.53 & 5.56 & 84 & 50 & 7.7 (0.8) \\
\enddata
\tablecomments{The numbers in parentheses indicate errors calculated from the absolute flux calibration error of 10\%.}
\end{deluxetable*}

\section{Discussion}

\subsection{Correlation among each species} \label{sec:d1}

\begin{figure*}[!th]
 \begin{center}
  \includegraphics[bb = 0 25 537 386, scale = 0.85]{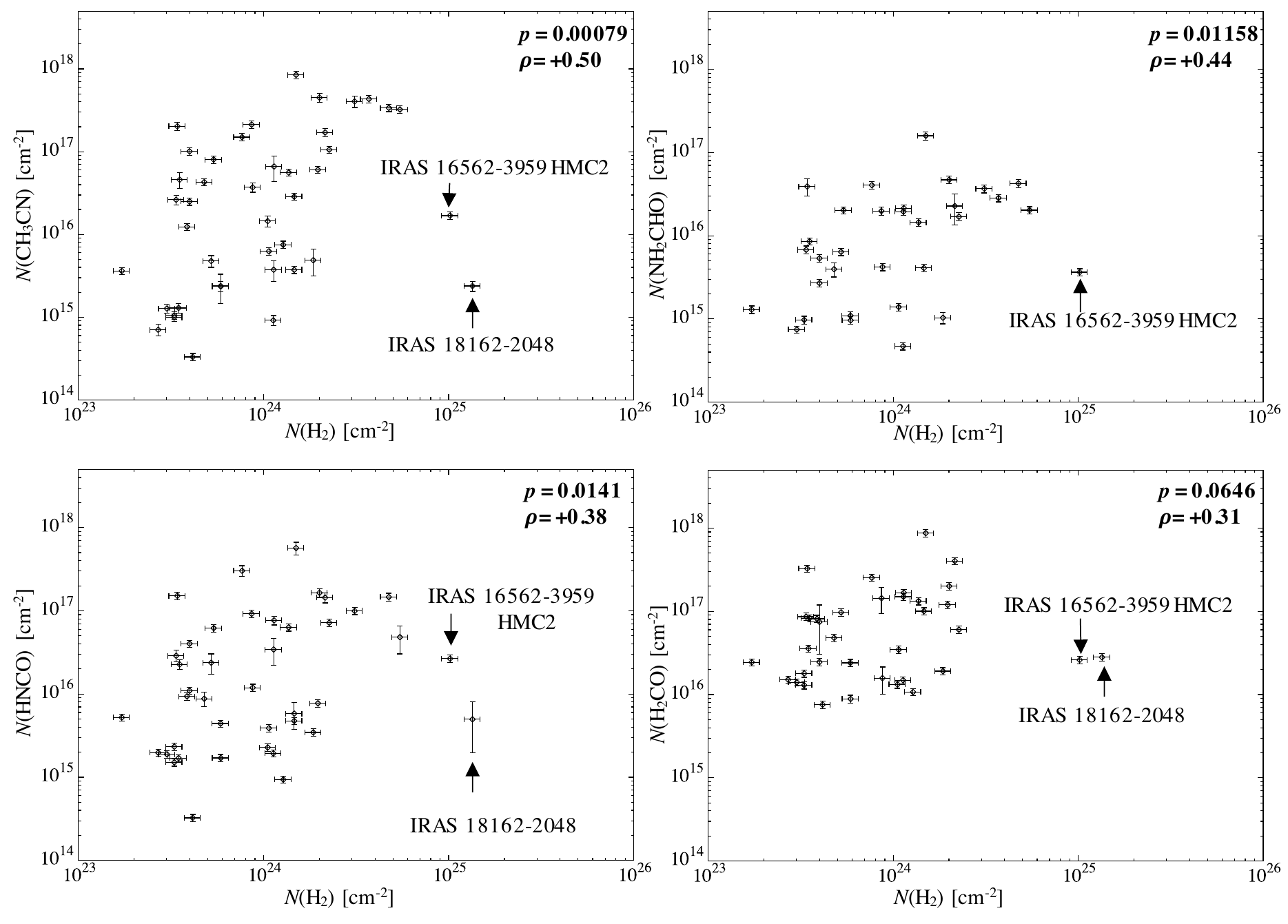}
 \end{center}
\caption{Relationships between the H$_{2}$ column density and molecular column densities. \label{fig:H2_mol}}
\end{figure*}

We search for correlations among the observed species to investigate their chemical links. 
As mentioned in Section \ref{sec:intro}, NH$_{2}$CHO has been considered to be closely chemically linked with HNCO \citep[e.g.,][]{2015MNRAS.449.2438L, 2019ESC.....3.2122L}.
In addition, chemical simulations suggest gas-phase formation of NH$_{2}$CHO from H$_{2}$CO in hot core regions \citep{2020ApJ...895...86G}.
On the other hand, CH$_{3}$CN is not expected to be directly related to the other observed species.

We use abundances derived as $X$($a$)=$N$($a$)/$N$(H$_{2}$), where $a$ represents molecular species, because comparisons using the column densities may mislead correlation coefficients.
If we investigate correlations using column densities, the total gas mass or total gas column density represented by $N$(H$_{2}$) could be the lurking third variable.
We made plots to investigate this possibility as shown in Figure \ref{fig:H2_mol}.
The outliers, labeled as `IRAS 16562-3959 HMC2' and `IRAS 18162-2048', show lower molecular column densities, nevertheless they have the highest H$_{2}$ column densities.
IRAS 16562-3959 HMC2 is an ultracompact \ion{H}{2} (UC \ion{H}{2}) region \citep{2018ApJS..236...45G, 2020ApJ...898...54T}. 
IRAS 18162-2048 possesses a highly collimated magnetized radio jet \citep{2010Sci...330.1209C}.
This jet may produce UV radiation via the strong shocks \citep{2017ApJ...847...58G}.
Thus, the lower molecular column densities in these two sources are likely caused by destruction of molecules by the UV radiation.

We conducted the Spearman's correlation coefficient test, and the derived $p$-values and correlation coefficients ($\rho$) are indicated in Figure \ref{fig:H2_mol}.
These statistical values mean that the molecular column densities have weak positive correlations with the H$_{2}$ column density.
We then decided to use the abundances to investigate chemical relationships among molecular species in the following parts of this paper.
Table \ref{tab:abundance} summarizes the derived abundances of each species, calculated from values summarized in Tables \ref{tab:mcmc} and \ref{tab:cont}.

\begin{deluxetable*}{llcccc}
\tabletypesize{\scriptsize}
\tablecaption{Molecular abundances \label{tab:abundance}}
\tablewidth{0pt}
\tablehead{
\colhead{Region} & \colhead{Position} & \colhead{$X$(CH$_{3}$CN)} & \colhead{$X$(NH$_{2}$CHO)} & \colhead{$X$(HNCO)} & \colhead{$X$(H$_{2}$CO)}
}
\startdata
G10.62-0.38 & HMC1 & ($4.7\pm0.7$)$\times10^{-8}$ & ($7.5\pm1.1$)$\times10^{-9}$ & ($3.2\pm0.4$)$\times10^{-8}$ & ($2.7\pm0.4$)$\times10^{-8}$	\\
& HMC2 & ($3.1\pm0.4$)$\times10^{-8}$ & ... & ($	3.9\pm0.6$)$\times10^{-9}$ &	($6.1\pm0.7$)$\times10^{-8}$ \\
G11.1-0.12 & HMC & ($3.2\pm0.5$)$\times10^{-9}$ & ($3.0\pm 0.4$)$\times10^{-9}$ & ($7.1\pm1.0$)$\times10^{-9}$ & ($3.9\pm0.6$)$\times10^{-8}$ \\
G11.92-0.61 & HMC & ($2.5\pm0.3$)$\times10^{-7}$ & ($2.3\pm0.3$)$\times10^{-8}$ & ($1.1\pm0.2$)$\times10^{-7}$ & ($1.7\pm0.6$)$\times10^{-7}$ \\
G14.22-0.50 S & HMC & ($8.1\pm1.4$)$\times10^{-10}$ & ($4.2\pm0.6$)$\times10^{-10}$ & ($1.7\pm0.2$)$\times10^{-9}$ & ($1.3\pm0.2$)$\times10^{-8}$ \\
G24.60+0.08 & HMC & ($2.6\pm0.5$)$\times10^{-9}$ & ... & ($7.3\pm1.0$)$\times10^{-9}$ & ($5.6\pm0.8$)$\times10^{-8}$ \\
G29.96-0.02 & HMC & ($5.7\pm0.8$)$\times10^{-7}$ & ($1.1\pm0.2$)$\times10^{-7}$ & ($3.8\pm0.8$)$\times10^{-7}$ & ($5.8\pm0.8$)$\times10^{-7}$ \\
G333.12-0.56 & HMC1 & ($6.3\pm0.9$)$\times10^{-8}$ & ($6.8\pm1.0$)$\times10^{-9}$ & ($2.8\pm0.4$)$\times10^{-8}$ & ($6.2\pm0.9$)$\times10^{-8}$ \\
& HMC2 & ($4.0\pm0.7$)$\times10^{-9}$ & ($1.6\pm0.2$)$\times10^{-9}$ & ($2.9\pm0.4$)$\times10^{-9}$ & ($1.5\pm0.2$)$\times10^{-8}$ \\
G333.23-0.06 & HMC1 & ($7.8\pm1.3$)$\times10^{-8}$ & ($2.0\pm0.3$)$\times10^{-8}$ & ($8.6\pm1.6$)$\times10^{-8}$ & ($2.6\pm0.4$)$\times10^{-7}$ \\
& HMC2 & ($2.6\pm1.0$)$\times10^{-9}$ & ($5.6\pm1.0$)$\times10^{-10}$ & ($1.9\pm0.3$)$\times10^{-9}$ & ($1.0\pm0.1$)$\times10^{-8}$ \\
G333.46-0.16 & HMC & ($5.9\pm0.8$)$\times10^{-9}$ & ... & ($7.3\pm1.0$)$\times10^{-10}$ & ($8.4\pm1.2$)$\times10^{-9}$ \\
& HMC1 & ($4.2\pm0.7$)$\times10^{-8}$ &	($4.8\pm0.7$)$\times10^{-9}$ & ($1.4\pm0.2$)$\times10^{-8}$ & ($1.8\pm0.7$)$\times10^{-8}$ \\
& HMC2 & ($2.1\pm0.3$)$\times10^{-8}$ &	($7.6\pm1.1$)$\times10^{-9}$ & ($3.0\pm0.4$)$\times10^{-8}$ & ($1.4\pm0.2$)$\times10^{-7}$ \\
G335.579-0.272 & HMC & ($1.3\pm0.2$)$\times10^{-7}$ & ($1.2\pm0.2$)$\times10^{-8}$ & ($3.2\pm0.5$)$\times10^{-8}$ & ... \\	
G335.78+0.17 & HMC1 & ($5.8\pm2.0$)$\times10^{-8}$ &	($1.9\pm0.3$)$\times10^{-8}$ & ($6.7\pm1.0$)$\times10^{-8}$ & ($1.5\pm0.2$)$\times10^{-7}$ \\
& HMC2 & ($1.3\pm0.3$)$\times10^{-7}$ & ($2.4\pm0.3$)$\times10^{-8}$ & ($6.4\pm1.1$)$\times10^{-8}$ & ($2.4\pm0.3$)$\times10^{-7}$ \\
G336.01-0.82 & HMC & ($1.5\pm0.2$)$\times10^{-7}$ & ($3.8\pm0.5$)$\times10^{-8}$ & ($1.2\pm0.2$)$\times10^{-7}$ & ... \\
G34.43+0.24 & HMC & ($7.1\pm1.0$)$\times10^{-8}$ & ($9.0\pm1.3$)$\times10^{-9}$ & ($3.1\pm0.5$)$\times10^{-8}$ & ...\\	
G34.43+0.24 MM2 & HMC & ($8.0\pm1.1$)$\times10^{-10}$ & ... & ($7.8\pm1.1$)$\times10^{-10}$ & ($1.8\pm0.3$)$\times10^{-8}$ \\
G35.03+0.35A & HMC & ($9.1\pm1.7$)$\times10^{-9}$ & ($1.2\pm0.2$)$\times10^{-8}$ & ($4.6\pm1.3$)$\times10^{-8}$ & ($1.9\pm0.3$)$\times10^{-7}$ \\
G35.13-0.74 & HMC1 & ($2.6\pm0.4$)$\times10^{-9}$ & ($2.8\pm0.4$)$\times10^{-9}$ & ($4.0\pm1.5$)$\times10^{-9}$ & ... \\
& HMC2 & ($4.1\pm1.6$)$\times10^{-9}$ & ($1.9\pm0.3$)$\times10^{-9}$ & ($7.5\pm1.1$)$\times10^{-9}$ & ($4.1\pm0.6$)$\times10^{-8}$ \\
& HMC3 & ($4.2\pm0.7$)$\times10^{-9}$ & ($2.5\pm0.4$)$\times10^{-9}$ & ($6.3\pm0.9$)$\times10^{-9}$ & ($4.7\pm0.7$)$\times10^{-8}$ \\
G35.20-0.74 N & HMC & ($4.1\pm0.6$)$\times10^{-8}$ & ($1.1\pm0.2$)$\times10^{-8}$ & ($4.6\pm0.7$)$\times10^{-8}$ & ($9.7\pm1.4$)$\times10^{-8}$ \\
G351.77-0.54 & HMC & ($1.2\pm0.3$)$\times10^{-7}$ & ($7.7\pm1.1$)$\times10^{-9}$ & ... & ... \\
G5.89-0.37 &	HMC1 & ($2.0\pm0.2$)$\times10^{-8}$ & ... & ($3.2\pm0.5$)$\times10^{-9}$ & ($6.9\pm1.0$)$\times10^{-8}$ \\
& HMC2 & ($3.2\pm0.5$)$\times10^{-8}$ & ... & ($2.4\pm0.3$)$\times10^{-8}$ & ($2.1\pm0.3$)$\times10^{-7}$ \\
G343.12-0.06 &HMC & ($8.0\pm1.2$)$\times10^{-8}$ & ($1.1\pm0.4$)$\times10^{-8}$ & ($6.7\pm1.1$)$\times10^{-8}$ & ($1.9\pm0.3$)$\times10^{-7}$ \\
IRAS 16562-3959 & HMC1 & ($2.5\pm0.4$)$\times10^{-7}$ & ($1.3\pm0.2$)$\times10^{-8}$ & ($1.0\pm0.1$)$\times10^{-7}$ & ($1.9\pm1.1$)$\times10^{-7}$ \\
& HMC2 & ($1.7\pm0.2$)$\times10^{-9}$ & ($3.6\pm0.5$)$\times10^{-10}$	& ($2.6\pm0.4$)$\times10^{-9}$ & ($2.6\pm0.4$)$\times10^{-9}$ \\
IRAS 18089-1732 & HMC1 & ($2.3\pm0.4$)$\times10^{-7}$ & ($2.3\pm0.3$)$\times10^{-8}$ & ($8.2\pm1.3$)$\times10^{-8}$ & ($1.0\pm0.1$)$\times10^{-7}$ \\
& HMC2 & ($5.8\pm0.8$)$\times10^{-9}$	& ($1.3\pm0.2$)$\times10^{-9}$ & ($3.6\pm0.5$)$\times10^{-9}$ & ($3.2\pm0.5$)$\times10^{-8}$ \\
IRAS 18151-1208	& HMC & ($3.7\pm0.6$$\times10^{-9}$ & ... & ($4.9\pm0.7$)$\times10^{-9}$ & ($1.0\pm0.2$)$\times10^{-7}$ \\
IRAS 18182-1433 & HMC1 & ($5.9\pm0.9$)$\times10^{-7}$ & ($1.1\pm0.3$)$\times10^{-7}$ & ($4.4\pm0.6$)$\times10^{-7}$ & ($9.5\pm1.4$)$\times10^{-7}$ \\
& HMC2 & ($9.0\pm1.3$)$\times10^{-8}$ & ($8.3\pm1.8$)$\times10^{-9}$ & ($1.8\pm0.4$)$\times10^{-8}$ & ($1.0\pm0.2$)$\times10^{-7}$ \\
& HMC3 & ($1.4\pm0.3$)$\times10^{-8}$ & ... & ($2.2\pm0.3$)$\times10^{-9}$ & ($1.3\pm0.2$)$\times10^{-8}$ \\
IRAS 18162-2048	& HMC & ($1.8\pm0.3$)$\times10^{-10}$ & ... & ($3.7\pm2.3$)$\times10^{-10}$ & ($2.1\pm0.3$)$\times10^{-9}$ \\
IRDC 18223-1243 & HMC & ($3.0\pm0.4$)$\times10^{-9}$ & ... & ($4.6\pm0.6$)$\times10^{-9}$ & ($5.4\pm0.8$)$\times10^{-8}$ \\					
NGC6334I N	& HMC1 & ($5.9\pm0.8$)$\times10^{-8}$ & ($3.7\pm0.5$)$\times10^{-9}$ & ($8.8\pm3.4$)$\times10^{-9}$ \\		
& HMC2 & ($3.3\pm1.0$)$\times10^{-9}$ & ($1.7\pm0.2$)$\times10^{-8}$ &	 ($3.0\pm1.1$)$\times10^{-8}$ & ($1.3\pm0.2$)$\times10^{-7}$ \\
W33A & HMC & ($2.0\pm0.2$)$\times10^{-7}$ & ($5.3\pm0.7$)$\times10^{-8}$ & ($4.0\pm0.6$)$\times10^{-7}$ & ($3.3\pm0.5$)$\times10^{-7}$ \\
\enddata
\tablecomments{Errors include the standard deviation derived from the MCMC analysis and the absolute flux calibration error of 10\%.}
\end{deluxetable*}

\begin{figure*}[th]
 \begin{center}
  \includegraphics[bb = 0 30 504 708, scale = 0.75]{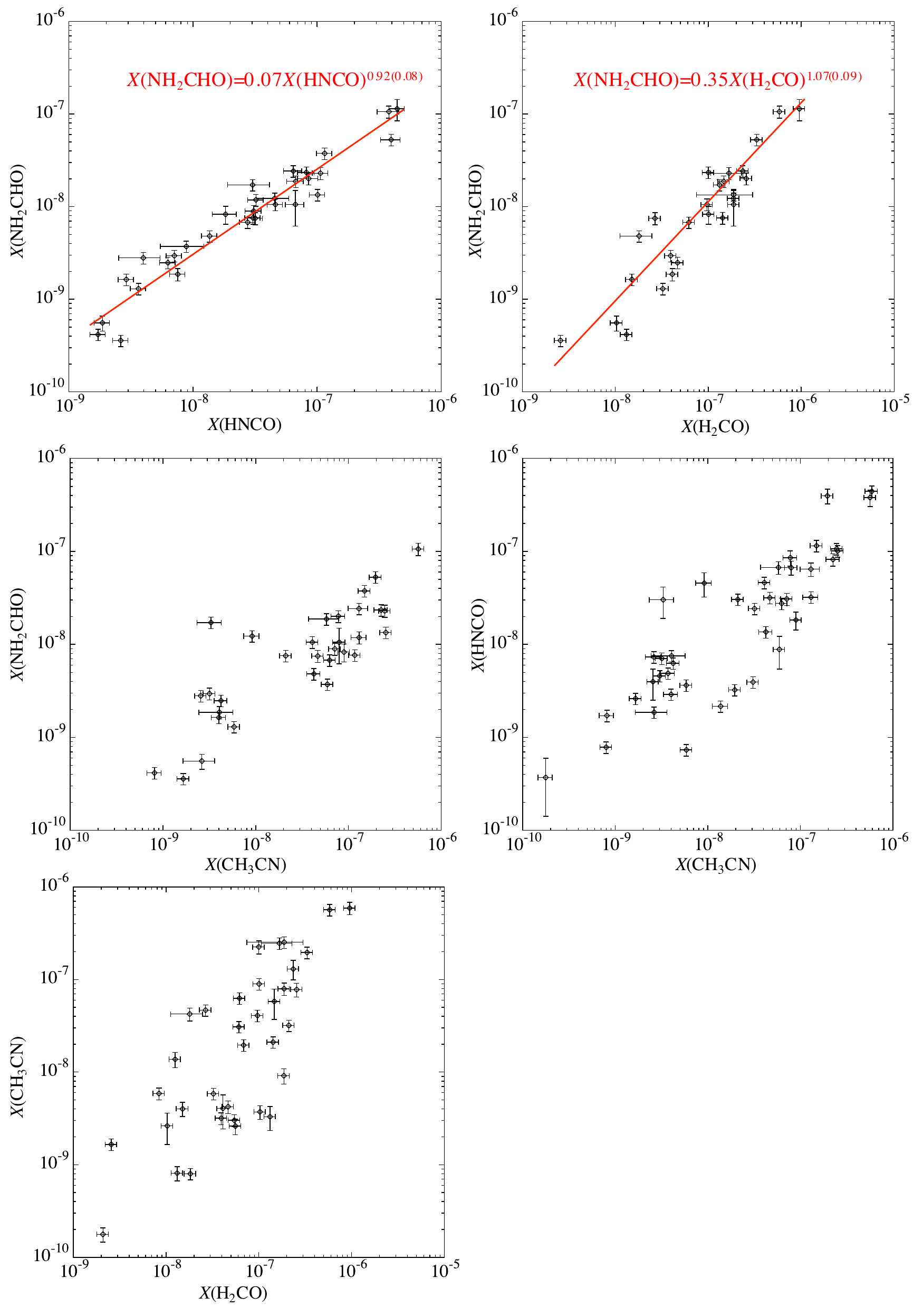}
 \end{center}
\caption{Relationships of molecular abundances. The red lines in the top two panels show the result of the best power-law fit. \label{fig:corr}}
\end{figure*}

Figure \ref{fig:corr} shows relationships between each pair of molecular species.
We conducted the Spearman's rank correlation test for each pair.
The  Spearman's correlation coefficients ($\rho$) are derived to be 0.96 and 0.91 for pairs of NH$_{2}$CHO-HNCO and NH$_{2}$CHO-H$_{2}$CO, respectively.
These high correlation coefficients imply that NH$_{2}$CHO may be chemically linked with HNCO and H$_{2}$CO.
These results are consistent with the previous studies and the prediction by chemical simulations \citep{2020ApJ...895...86G}.

We fitted the plot of NH$_{2}$CHO vs. HNCO, as \citet{2015MNRAS.449.2438L} analyzed.
The best power-law fit is $X$(NH$_{2}$CHO)=0.07$X$(HNCO)$^{0.92(\pm0.08)}$.
\citet{2015MNRAS.449.2438L} derived the best power-law fit as $X$(NH$_{2}$CHO)=0.04$X$(HNCO)$^{0.93}$, using data toward YSOs with various stellar masses, from low- through intermediate- to high-mass stars.
Our result agrees with that derived by \citet{2015MNRAS.449.2438L} well.
Our data set contains a sample of targets with larger abundances of NH$_{2}$CHO ($\approx3\times10^{-10}-10^{-7}$) than those in \citet{2015MNRAS.449.2438L} ($\approx3\times10^{-11}-10^{-8}$), and our sample is the largest one in high-mass star-forming regions, to our best knowledge.
The same best power-law fits for different ranges of abundances likely mean that the chemistry (formation and destruction processes) of NH$_{2}$CHO is common around YSOs with various stellar masses.

Since we found the tight correlation between NH$_{2}$CHO and H$_{2}$CO, we also fitted the plot of NH$_{2}$CHO vs. H$_{2}$CO with the same method.
The best power-law fit is $X$(NH$_{2}$CHO)=0.35$X$(H$_{2}$CO)$^{1.07(\pm0.09)}$.

The second and third rows of Figure \ref{fig:corr} show correlations between CH$_{3}$CN and NH$_{2}$CHO/HNCO/H$_{2}$CO.
Although there are no expected direct chemical relationships between CH$_{3}$CN and the other species, there is a possibility that CH$_{3}$CN is positively correlated with the other species.
The plots show more scatter compared to the two plots in the first row, suggestive of weaker correlations. 
The Spearman's correlation coefficients are derived to be 0.83, 0.85, 0.74 for pairs of CH$_{3}$CN-NH$_{2}$CHO, CH$_{3}$CN-HNCO, and CH$_{3}$CN-H$_{2}$CO, respectively.
These correlation coefficients are lower than those for pairs NH$_{2}$CHO-HNCO and NH$_{2}$CHO-H$_{2}$CO, but they indicate still strong correlations ($\rho \geq 0.7$).
These correlations may mean that all of the species react to temperature in the same manner rather than chemical links \citep{2018MNRAS.474.2796Q}.
To explore this point further, we will consider a partial correlation test in the next subsection.

\subsection{Partial correlation test} \label{sec:d2}

\begin{figure*}[th]
 \begin{center}
  \includegraphics[bb = 0 30 532 527, scale = 0.7]{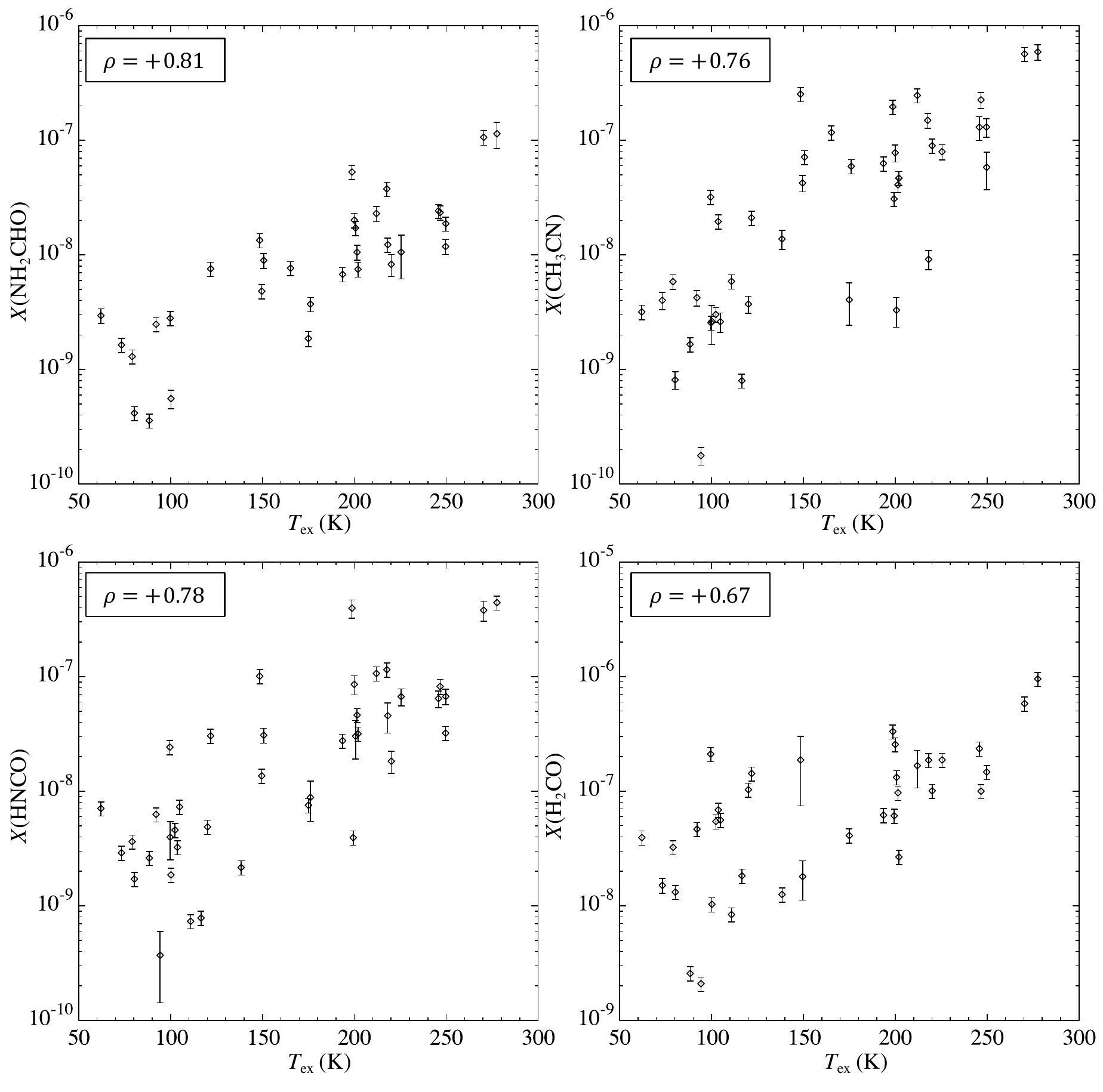}
 \end{center}
\caption{Relationships between excitation temperatures and molecular abundances. \label{fig:corrTex}}
\end{figure*}

As we derived in Section \ref{sec:d1}, all of the species show positive correlations.
However, as \citet{2018MNRAS.474.2796Q} pointed out, the observed correlations may not indicate chemical links, but they could reflect that all of the species act similarly against temperature.
For example, assuming that molecules form on dust surfaces in the cold starless core stage and sublimate into the gas phase in the hot core regions with temperatures above 100 K, their abundances increase as the temperature increases.
In this subsection, we will conduct a partial correlation test to investigate pure chemical links excluding the effect of temperature.

Figure \ref{fig:corrTex} shows relationships between excitation temperatures and molecular abundances.
We found that all of the molecular abundances show positive correlations with the excitation temperatures; $\rho$ values are 0.81, 0.78, 0.67, and 0.76 for NH$_{2}$CHO, HNCO, H$_{2}$CO, and CH$_{3}$CN, respectively.
Hence, the correlation among each species derived in Section \ref{sec:d1} may be fake, and the temperature may act as the lurking third variable.

In order to derive the pure chemical links excluding the lurking third variable, or the temperature, we conducted a partial correlation test.
Assuming local thermodynamics equilibrium (LTE), we use the excitation temperature derived from CH$_{3}$CN analyses to represent the gas kinetic temperature and dust temperature. 
The partial correlation test removes the mutual dependence of the two first variables on the third one.
We utilize the excitation temperature derived from the analyses of the CH$_{3}$CN ($^{13}$CH$_{3}$CN) lines as a representative temperature.
We adopted the first-order partial correlation coefficient \citep[e.g.,][]{2012psa..book.....W,2018MNRAS.473.1059U}:
\begin{equation}
\rho_{12,3} = \frac{\rho_{12}-\rho_{13}\rho_{23}}{\sqrt{(1-\rho^{2}_{13})(1-\rho^{2}_{23})}}.
\end{equation}
Here, we are interested in molecular abundances (1 and 2), and the temperature is considered as the third variable (3).

\begin{deluxetable*}{lccccc}
\tablecaption{Summary of partial correlation coefficient \label{tab:partialtest}}
\tablewidth{0pt}
\tablehead{
\colhead{} & \colhead{HNCO vs. NH$_{2}$CHO} & \colhead{H$_{2}$CO vs. NH$_{2}$CHO} & \colhead{CH$_{3}$CN vs. HNCO} & \colhead{CH$_{3}$CN vs. NH$_{2}$CHO} & \colhead{H$_{2}$CO vs. CH$_{3}$CN}
}
\startdata
$\rho_{12}$ & 0.96 & 0.91 & 0.85 & 0.83 & 0.74\\
$\rho_{13}$ & 0.78 & 0.67 & 0.76 & 0.76 & 0.67 \\
$\rho_{23}$ & 0.81 & 0.81 & 0.78 & 0.81 & 0.76 \\
\cline{1-6}
{\bf {$\rho_{12,3}$}} & {\bf {0.89}} & {\bf {0.84}} & {\bf {0.64}} & {\bf {0.55}} & {\bf {0.48}} \\
\enddata
\tablecomments{``1'' and ``2'' correspond to Species A and Species B, respectively (Species A vs. Species B written at the top of table). ``3'' corresponds to the excitation temperature.}
\end{deluxetable*}

Table \ref{tab:partialtest} summarizes all of the correlation coefficients for each pair of molecules.
The derived partial correlation coefficients are summarized in the last row.

The partial correlation coefficients for HNCO-NH$_{2}$CHO and H$_{2}$CO-NH$_{2}$CHO are still high ($>0.8$), whereas those for the other pairs with CH$_{3}$CN become lower ($<0.65$).
These results most likely indicate that NH$_{2}$CHO is chemically linked with HNCO and H$_{2}$CO.
Thus, the observed correlation between NH$_{2}$CHO and HNCO is not the result of having the same response of the two species to temperature, as pointed out by \citet{2018MNRAS.474.2796Q}.
On the other hand, CH$_{3}$CN is less chemically linked with the other species studied here.
We found that the partial correlation test is a strong tool to statistically investigate chemical relationships.

We further investigated the correlation coefficients between the molecular abundances and excitation temperatures.
The highest correlation coefficient is derived for NH$_{2}$CHO (0.81), followed by HNCO (0.78), CH$_{3}$CN (0.76), and H$_{2}$CO (0.67). 
This order matches with the order of the calculated binding energies of each species (Table \ref{tab:molline}).
This also suggests that thermal desorption is important for these molecules.
Since the emission of a molecule with a high binding energy should be dominated by the central hot core region, the correlation with the temperature should be strong.
Hence, the derived correlation coefficients with the temperature suggest that the NH$_{2}$CHO emission comes from the central hot core region, whereas the emission of H$_{2}$CO could come not only from the inner hot core but also from the outer part of cores.
This is consistent with the spatial distribution of the molecular lines (Figures \ref{fig:mom01} -- \ref{fig:mom04}), as mentioned in Section \ref{sec:mom0}.

\subsection{Comparison with chemical models} \label{sec:d3}

We compare the observed molecular abundances with the chemical models developed by \citet{2020ApJ...895...86G}.
We consider a linear density variation with a slope $\frac{\rho_{\rm {max}} - \rho_{\rm {min}}}{t_{\rm {f}}-t_{\rm {i}}}$, where $t_{\rm {f}}-t_{\rm {i}}$ corresponds to the collapse timescale.
Similarly, the temperature increases with a linear slope during the warm-up stage.

The dual-cyclic hydrogen addition and abstraction reactions \citep{Haupa...2019} which occur in the gas phase are included in the models.
This cycle consists of four reactions.
From HNCO to NH$_{2}$CHO, two hydrogen addition reactions occur:
\begin{equation} \label{equ:Ha1}
{\rm {HNCO}} + {\rm {H}} \rightarrow  {\rm {H}}_2{\rm {NCO}},
\end{equation} 
followed by
\begin{equation} \label{equ:Ha2}
{\rm {H}}_2{\rm {NCO}} + {\rm {H}} \rightarrow  {\rm {NH}}_2{\rm {CHO}}.
\end{equation} 
In a sequence from NH$_{2}$CHO to HNCO, two hydrogen abstraction reactions happen:
\begin{equation} \label{equ:Ha3}
{\rm {NH}}_2{\rm {CHO}} + {\rm {H}} \rightarrow  {\rm {H}}_2{\rm {NCO}} + {\rm {H}}_{2},
\end{equation} 
followed by
\begin{equation} \label{equ:Ha4}
{\rm {H}}_2{\rm {NCO}} + {\rm {H}} \rightarrow  {\rm {HNCO}}+ {\rm {H}}_{2}.
\end{equation} 
Reactions (\ref{equ:Ha2}) and (\ref{equ:Ha4}) are barrierless, whereas Reactions (\ref{equ:Ha1}) and (\ref{equ:Ha3}) have activation barriers of 2530 -- 5050 K and 240 -- 3130 K, respectively. 
In the models of \citet{2020ApJ...895...86G}, the gas-phase reaction between NH$_{2}$ and H$_{2}$CO to form NH$_{2}$CHO is included:
\begin{equation} \label{equ:H2CO}
{\rm {NH}}_2+{\rm {H}}_2{\rm {CO}} \rightarrow  {\rm {NH}}_2{\rm {CHO}} + {\rm {H}}.
\end{equation}

The hydrogenation reaction leads to the formation of CH$_{3}$CN on dust surfaces in the cold phase. 
In contrast, its formation by a dust-surface radical-radical reaction between CN and CH$_{3}$ is efficient during the initial warm-up phase.
In addition, CH$_{3}$CN could form by the following barrierless reaction on dust surfaces:
\begin{equation} \label{eq:CH3CN1}
{\rm {H}} + {\rm{H}}_2{\rm{CCN}} \rightarrow {\rm {CH}}_3{\rm{CN}}.
\end{equation}
The major gas-phase formation route of CH$_{3}$CN is the dissociative recombination reaction of CH$_{3}$CNH$^{+}$.
Destruction routes of CH$_{3}$CN on dust surfaces are cosmic-ray induced desorption, photodissociation, and thermal and non-thermal desorption.
Furthermore, the models of \citet{2020ApJ...895...86G} include the successive hydrogenation reactions of CH$_{3}$CN to lead finally to ethanimine (CH$_{3}$CHNH). 

There are six physical parameters in the models of \citet{2020ApJ...895...86G}; the maximum density ($\rho_{\rm {max}}$ [cm$^{-3}$]), maximum temperature ($T_{\rm {max}}$ [K]), collapsing timescale ($t_{\rm {coll}}$ [yr]), warm-up timescale ($t_{\rm {w}}$ [yr]), post-warm-up timescale ($t_{\rm {pw}}$ [yr]), and initial dust temperature ($T_{\rm {ice}}$ [K]).
\citet{2020ApJ...895...86G} ran two types of models, Model A and Model B, as summarized in Table 9 in their paper.
In Model A, the physical parameters are fixed with the following values; $\rho_{\rm {max}}=10^7$ cm$^{-3}$, $T_{\rm {max}} = 200$ K, $t_{\rm {coll}}=10^6$ yr, $t_{\rm {w}}=5\times10^4$ yr, $t_{\rm {pw}}$ = (6.2--10)$\times10^{4}$ yr, and $T_{\rm {ice}}=20$ K.
For Model B, $\rho_{\rm {max}}$, $T_{\rm {max}}$, and $t_{\rm {coll}}$ varied in ranges of $10^5-10^7$ cm$^{-3}$, $100-400$ K, and $10^5-10^6$ yr, respectively.
The warm-up timescale, post-warm-up timescale, and initial dust temperatures were fixed at $5\times10^4$ yr, $10^5$ yr, and 20 K, respectively, in Model B.

\begin{deluxetable*}{lcccc}
\tablecaption{Summary of representative observed abundances \label{tab:repabundance}}
\tablewidth{0pt}
\tablehead{
\colhead{} & \colhead{HNCO} & \colhead{NH$_{2}$CHO} & \colhead{H$_{2}$CO} & \colhead{CH$_{3}$CN}
}
\startdata
Minimum (Min) & $3.70\times10^{-10}$ & $3.59\times10^{-10}$ & $2.09\times10^{-9}$ & $1.77\times10^{-10}$ \\
Maximum (Max) & $4.42\times10^{-7}$ & $1.14\times10^{-7}$ & $9.54\times10^{-7}$ &  $5.90\times10^{-7}$ \\
Average (Ave.) & $5.64\times10^{-8}$ & $1.77\times10^{-8}$ & $1.31\times10^{-7}$ & $8.11\times10^{-8}$ \\
Median (Med.) & $2.13\times10^{-8}$ & $8.61\times10^{-9}$ & $6.52\times10^{-8}$ & $3.13\times10^{-8}$  \\
\enddata
\end{deluxetable*}

\begin{figure}[!th]
 \begin{center}
  \includegraphics[bb = 10 20 302 610, scale = 0.8]{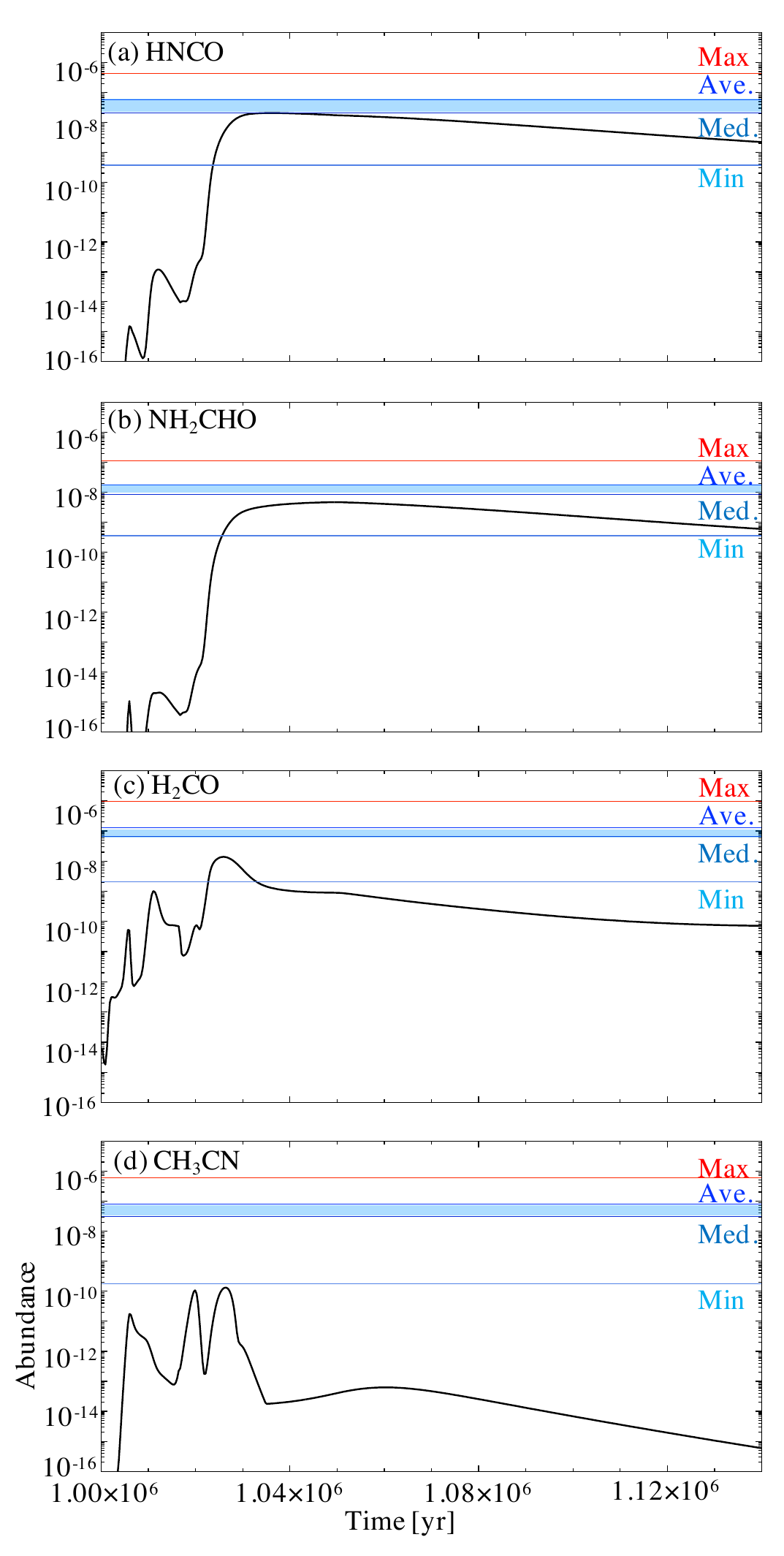}
 \end{center}
\caption{Comparison with Model A \citep{2020ApJ...895...86G}. Panels (a) -- (d) show results of HNCO, NH$_{2}$CHO, H$_{2}$CO, and CH$_{3}$CN, respectively. Black curves indicate the modeled abundances (gas phase). The four representative observed abundances are plotted (Maximum, Average, Median, and Minimum). The blue filled ranges indicate the ranges between average and median values. \label{fig:modelA}}
\end{figure}

Figure \ref{fig:modelA} shows comparisons between the observed abundances and the modeled abundances obtained by Model A \citep{2020ApJ...895...86G}.
As the observed values, we plot the four representative values from the whole hot core populations; maximum, average, median, and minimum values (Table \ref{tab:repabundance}).

The minimum abundances agree with the model around $1.02\times10^6$ yr ($T\approx100$ K).
The minimum observed CH$_{3}$CN abundance agrees with model at the almost same age within a factor of 1.3.
The minimum abundance of NH$_{2}$CHO is derived at IRAS\,16562-3959 HMC2, and those of HNCO and H$_{2}$CO are derived in IRAS\,18162-2048.
The derived excitation temperatures of CH$_{3}$CN in these HMCs are 88 K and 94.3 K, which are lower than the other sources and close to the modeled value of 100 K.
In addition, in the case of IRAS\,18162-2048, as mentioned before, the ionized jet may explain the lower molecular column densities.

The observed average and median abundances of HNCO and NH$_{2}$CHO are most close to the modeled values around ($\approx1.04-1.05$)$\times10^{6}$ yr ($T\approx165-200$ K).
The temperature ranges which can reproduce the observed average and median values agree with the average excitation temperature at the whole hot core populations ($\approx162$ K).
Thus, the tendencies of the observed abundances for both HNCO and NH$_{2}$CHO agree with the modeled results.
The median value of H$_{2}$CO is close to the modeled abundance around $\approx1.03\times10^{6}$ yr, which is almost consistent with the ages constrained by HNCO and NH$_{2}$CHO (($\approx1.04-1.05$)$\times10^{6}$ yr), but in general, the modeled H$_{2}$CO abundances tend to be lower than the observed ones.
The observed maximum abundances cannot be explained by Model A, implying that assumed physical parameters are not suitable for the hot core with the maximum molecular abundances.

\begin{figure*}[!th]
 \begin{center}
  \includegraphics[bb = 0 20 545 460, scale = 1.0]{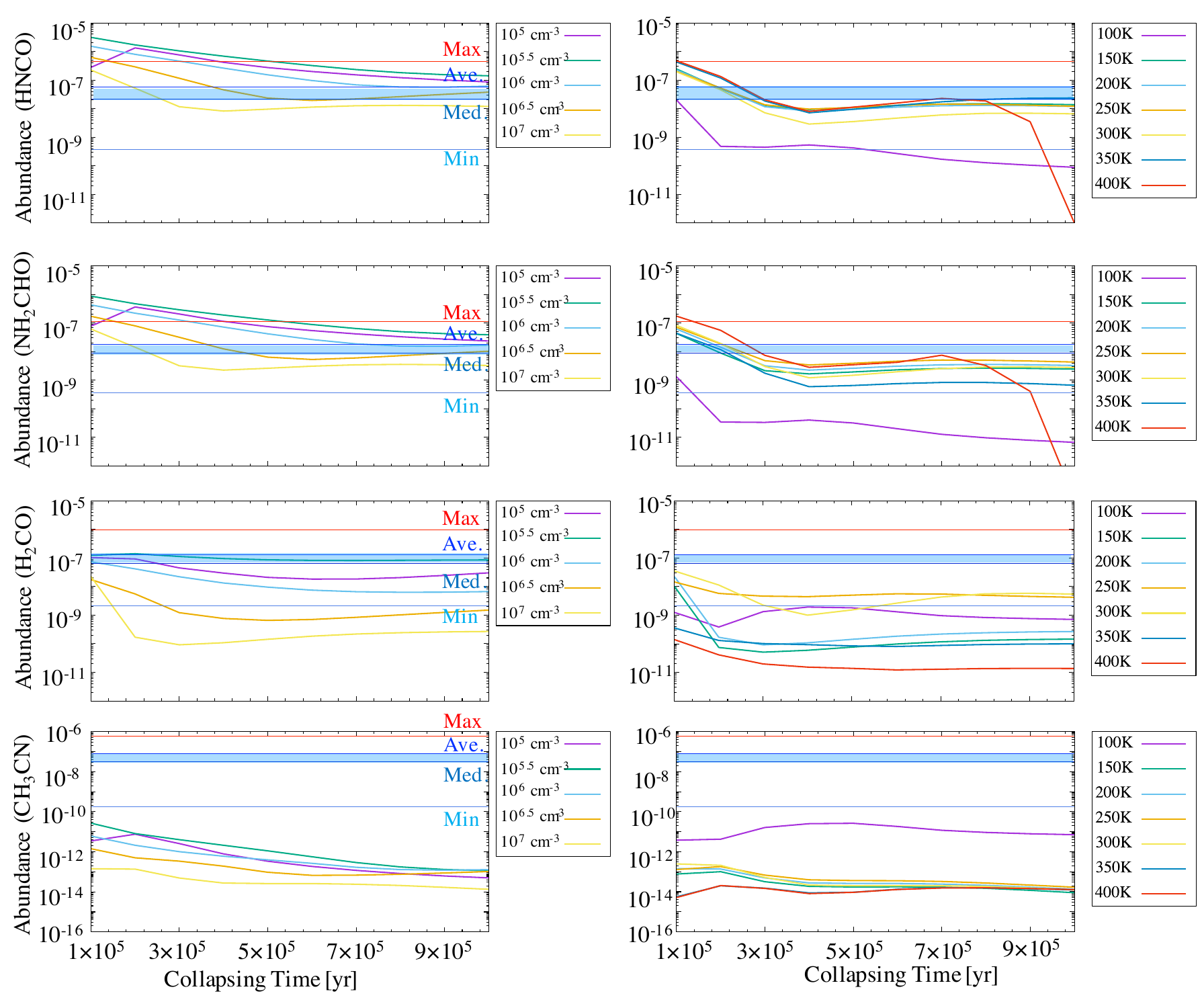}
 \end{center}
\caption{Comparison with Model B \citep{2020ApJ...895...86G}. Panels from top to bottom show results of HNCO, NH$_{2}$CHO, H$_{2}$CO, and CH$_{3}$CN, respectively. The modeled abundances plotted here are the values at the end of the simulations; $t$ = $t_{\rm {collapse}}$ + $t_{\rm {warm-up}}$ ($5\times10^4$ yrs)+ $t_{\rm {post\,warm-up}}$ ($10^5$ yrs). Left panels show dependences on different collapsing timescale and maximum density with the maximum temperature of 200 K. Right panels show dependences on different collapsing timescale and maximum temperature with the maximum density of $10^7$ cm$^{-3}$. The four representative observed abundances are plotted (Maximum, Average, Median, and Minimum). The blue filled ranges indicate the ranges between average and median values. \label{fig:modelB}}
\end{figure*}

Figure \ref{fig:modelB} shows a comparison between the representative observed abundances and modeled abundances obtained by Model B \citep{2020ApJ...895...86G}.
Left panels show dependences on different $t_{\rm {coll}}$ and $\rho_{\rm {max}}$, and right panels indicate dependences on different $t_{\rm {coll}}$ and $T_{\rm {max}}$, respectively.
The initial dust temperature ($T_{\rm {ice}}$) is fixed at 20 K, and $t_{\rm {pw}}$ is $10^{5}$ yr.
The modeled abundances plotted here are the values at the end of the simulations; $t$ = $t_{\rm {collapse}}$ + $t_{\rm {warm-up}}$ ($5\times10^4$ yrs)+ $t_{\rm {post\,warm-up}}$ ($10^5$ yrs).
As seen in Figure \ref{fig:modelB}, the maximum observed abundances could not be reproduced by the models in most cases, because the plotted modeled abundances are the values at the end of the simulation.
We show the modeled maximum abundances obtained by Model B in Figure \ref{fig:modelBpeak} in Appendix \ref{sec:a3}.
The modeled maximum abundances with some cases consist with the observed maximum abundances by less than one order of magnitude.

In Figure \ref{fig:modelB}, we found low abundances of H$_{2}$CO and high abundances of NH$_{2}$CHO in the models.
We attribute these results to the uncertainty of the reaction rate coefficient for the gas-phase reaction between NH$_{2}$ and H$_{2}$CO (Reaction (\ref{equ:H2CO})).
We used the $\alpha$ value of $5.00\times10^{-12}$, but \citet{2017MNRAS.468L...1S} derived the value at $7.79\times10^{-15}$ by their quantum chemical calculation.
However, the $\gamma$ value provided by \citet{2017MNRAS.468L...1S} was $\approx5.5$ times lower than that of \citet{2015MNRAS.453L..31B}.
Here, we used the $\gamma$ value from \citet{2017MNRAS.468L...1S}.
If the $\alpha$ value for this reaction becomes lower, as suggested by \citet{2017MNRAS.468L...1S}, the conversion rate from H$_{2}$CO to NH$_{2}$CHO should be slow.
In addition to the $\alpha$, $\beta$ and $\gamma$ values used in \citet{2020ApJ...895...86G} for this reaction, we have also evaluated the modeling results considering the $\alpha$, $\beta$ and $\gamma$ values available in \citet{2017MNRAS.468L...1S}. 
We did not find significant changes for our best-fit results that are obtained comparing with observation. 
Only a minor change in collapsing time has been noticed.

The maximum abundances of HNCO and NH$_{2}$CHO can be reproduced by the models with $\rho_{\rm {max}}=10^{5}-10^{6.5}$ cm$^{-3}$, as seen in the left panels. 
The average and median abundances of these two species agree with the following three models; (1) $\rho_{\rm {max}}=10^{6}$ cm$^{-3}$ and $t_{\rm {coll}}\approx8\times10^{5}$ yr, (2) $\rho_{\rm {max}}=10^{6.5}$ cm$^{-3}$ and $t_{\rm {coll}}=4\times10^{5}$ yr, and (3) $\rho_{\rm {max}}=10^{7}$ cm$^{-3}$ and $t_{\rm {coll}}=2\times10^{5}$ yr.
The models with higher $\rho_{\rm {max}}$ values need to collapse rapidly to reproduce the observed abundances.
Their minimum abundances cannot be reproduced by the models, because the adopted maximum temperature (200 K) is too high for these sources (88--95 K; see in the previous paragraph).
In fact, the minimum abundances can be reproduced by the model with $T_{\rm {max}}=100$ K in the right panels, as discussed later.

The minimum H$_{2}$CO abundance is reproduced by the models with $\rho_{\rm {max}}=10^{6.5}-10^{7}$ cm$^{-3}$, while the average and median values agree with the model with $\rho_{\rm {max}}=10^{5.5}$ cm$^{-3}$. 
In addition, the models with $\rho_{\rm {max}}=10^{5}$ cm$^{-3}$ and $10^{6}$ cm$^{-3}$ reproduce the average and median abundances around $t_{\rm {coll}}\approx$(1--2)$\times10^{5}$ yr.
The observational results of H$_{2}$CO tend to show agreements with lower-$\rho_{\rm {max}}$ models compared to the cases of HNCO and NH$_{2}$CHO ($\rho_{\rm {max}}=10^{6}-10^{7}$ cm$^{-3}$).
These results would be consistent with the fact that the H$_{2}$CO emission shows more spatially extended features than the other molecules (Figures \ref{fig:mom01}--\ref{fig:mom04} in Appendix \ref{sec:a1}).
The maximum value, derived in IRAS\,18182-1433 HMC1, has not been reproduced by the models in Figure \ref{fig:modelB}, but the modeled maximum abundances with $\rho_{\rm {max}}=10^{5}$ cm$^{-3}$ agree with the observed maximum value within one order of magnitude (see Figure \ref{fig:modelBpeak} in Appendix \ref{sec:a3}).

As seen in the right panels of HNCO and NH$_{2}$CHO in Figure \ref{fig:modelB}, the maximum abundances prefer shorter collapsing time with highest $T_{\rm {max}}$ of 400 K.
The maximum abundances of the both molecules are derived in IRAS\,18182-1433 HMC1.
The excitation temperature at HMC1 is the highest value ($\approx 278$ K) among the observed hot cores.
This is consistent with the agreement of the maximum abundance with the highest $T_{\rm {max}}$ model.
On the other hand, HMC2 and HMC3 in the same region do not show high abundances and NH$_{2}$CHO has not been detected at HMC3.
Hence, in this high-mass star-forming region, it is likely that an MYSO in HMC1 was firstly born rapidly, and the other sources were slowly formed later. 

The average and median values of HNCO and NH$_{2}$CHO are consistent with the models with $T_{\rm {max}}=150-400$ K.
The best agreed conditions are $t_{\rm {coll}}=$(2--3)$\times10^{5}$ yr, and the models with lower $T_{\rm {max}}$ values need to collapse quickly.

In the case of H$_{2}$CO, the minimum observed abundance agrees with the models within one order of magnitude except for $T_{\rm {max}}=400$ K, but the other three representative values disagree with the models.
This could be explained by the fact that the observed H$_{2}$CO abundances prefer the models with low-$\rho_{\rm {max}}$ values, while the right panels show results with $\rho_{\rm {max}}=10^{7}$ cm$^{-3}$.

The observed CH$_{3}$CN abundances disagree in most cases, but CH$_{3}$CN does not take parts in the formation network of NH$_{2}$CHO. 
This disagreement does not undermine our analysis for NH$_{2}$CHO.   
The minimum value agrees with the models of $T_{\rm {max}}=100$ K and $t_{\rm {coll}}\approx$(3-6)$\times10^{5}$ yr within one order of magnitude.
As seen in Figure \ref{fig:modelBpeak} in Appendix \ref{sec:a3}, the observed CH$_{3}$CN abundances agree with the modeled maximum abundances with low-$\rho_{\rm {max}}$ models ($10^5-10^{5.5}$ cm$^{-3}$), not at end of the simulations.
This could mean that CH$_{3}$CN is deficient at the age of the end of the simulations due to its destruction, and the CH$_{3}$CN gas may trace chemically younger gas around MYSOs.

We note that there are still uncertainties in reaction rate constants for the dual-cyclic hydrogen addition and abstraction reactions.
We tested changing the $\alpha$ value by two orders of magnitude (from $\alpha=10^{-10}$ to $\alpha=10^{-12}$).
The modeled HNCO abundance becomes lower by three orders of magnitude, and the NH$_{2}$CHO abundance increase by one order of magnitude.
Hence, the modeled HNCO abundance with $\alpha=10^{-12}$ cannot explain the observed abundances, and the rate constants used by \citet{2020ApJ...895...86G}, $\alpha=10^{-10}$, seem to be more reasonable.
However, measurements to estimate the rate constants of the reactions involved are necessary to better understand relationships among these species in various astronomical environments.

In summary, most of the sources observed by the DIHCA project are at the hot core stage with temperatures of 150--400 K and densities of $10^{6}-10^{7}$ cm$^{-3}$, judging from HNCO and NH$_{2}$CHO.
These values are consistent with typical hot core values.
On the other hand, the observed H$_{2}$CO and CH$_{3}$CN abundances prefer models with lower density conditions of $\sim10^{5}-10^{5.5}$ cm$^{-3}$.
This is consistent with the spatial distribution of the observed emission.
Thus, the contribution from the outer parts of cores is larger in the case of H$_{2}$CO and CH$_{3}$CN, while NH$_{2}$CHO and HNCO emission comes from inner central cores.
These are consistent with the expectation based on the calculated binding energy as discussed in Section \ref{sec:d2}.

\section{Conclusions} \label{sec:conclusion}

We have analyzed molecular lines of NH$_{2}$CHO, HNCO, H$_{2}$CO, and CH$_{3}$CN ($^{13}$CH$_{3}$CN) toward the 30 high-mass star-forming regions targeted by the DIHCA project.
The angular resolutions of $\sim0\farcs3$ spatially resolve hot molecular cores (HMCs).
The main conclusions of this paper are as follows.

\begin{enumerate}
\item The lines of CH$_{3}$CN, HNCO, and H$_{2}$CO have been detected from 29 high-mass star-forming regions (96.7\%), and the lines of NH$_{2}$CHO have been detected from 23 regions (76.7\%).
Thanks to a large number of detections of the target species, we statistically investigated their chemical links.
A total of 44 HMCs have been identified in the moment 0 maps of CH$_{3}$CN.

\item The identified cores have the excitation temperatures of $\sim62-278$ K and H$_{2}$ column densities of $1.7\times10^{23}-2.9\times10^{25}$ cm$^{-2}$, respectively. We have derived molecular abundances with respect to H$_{2}$.

\item We have investigated correlations between NH$_{2}$CHO and HNCO and  between NH$_{2}$CHO and H$_{2}$CO by applying a partial correlation test in order to investigate pure chemical links excluding a possible lurking third variable (temperature in this case).
The derived correlation coefficients are 0.89 and 0.84 for pairs of NH$_{2}$CHO-HNCO and NH$_{2}$CHO-H$_{2}$CO, respectively.
These strong correlations indicate that they are most likely chemically linked in hot cores.

\item We have fitted the abundance plots of HNCO vs. NH$_{2}$CHO and H$_{2}$CO vs. NH$_{2}$CHO.
We have obtained the best power-law fit of $X$(NH$_{2}$CHO) = 0.07$X$(HNCO)$^{0.92}$, which is well consistent with a previous study \citep[$X$(NH$_{2}$CHO) = 0.04$X$(HNCO)$^{0.93}$;][]{2015MNRAS.449.2438L}.
Their abundances studied in this paper are higher than those of previous studies, and then we have confirmed that this relationship is applicable in a wide range of their abundances.
The same power-law fit from low- to high-mass star-forming regions suggest that chemistry of NH$_{2}$CHO is common around YSOs with various stellar masses.
The best power-law fit for the plot of H$_{2}$CO vs. NH$_{2}$CHO is $X$(NH$_{2}$CHO)=0.35$X$(H$_{2}$CO)$^{1.07}$.

\item We have compared the observed abundances and chemical models including the dual-cyclic hydrogen addition and abstraction reactions between HNCO and NH$_{2}$CHO and the gas-phase formation of NH$_{2}$CHO by the reaction between NH$_{2}$ and H$_{2}$CO.
The models can reproduce the observed molecular abundances.
The observed abundances of HNCO and NH$_{2}$CHO prefer the models with temperatures of 150--400 K and densities of $10^{6}-10^{7}$ cm$^{-3}$, which agree with typical hot core values.
On the other hand, the H$_{2}$CO and CH$_{3}$CN abundances prefer the models with lower maximum densities ($\sim10^{5}-10^{5.5}$ cm$^{-3}$).
These results mean that the HNCO and NH$_{2}$CHO emission comes from inner cores, whereas the contributions from the outer part of cores are mixed in the case of the H$_{2}$CO and CH$_{3}$CN emission.
This scenario is consistent with the spatial distributions of each species (the H$_{2}$CO and CH$_{3}$CN emission is more extended than the other two species) and the calculated binding energies.
\end{enumerate}

\begin{acknowledgments}
This paper makes use of the following ALMA data: ADS/JAO.ALMA\#2016.1.01036.S and 2017.1.00237.S. 
ALMA is a partnership of ESO (representing its member states), NSF (USA), and NINS (Japan), together with NRC (Canada), MOST and ASIAA (Taiwan), and KASI (Republic of Korea), in cooperation with the Republic of Chile. The Joint ALMA Observatory is operated by ESO, AUI/NRAO, and NAOJ.
K.T. is supported by JSPS KAKENHI grant No. JP20K14523.
P.S. was financially supported by Grant-in-Aid for Scientific Research (KAKENHI Number  22H01271) of Japan Society for the Promotion of Science (JSPS).
P. G acknowledges the support from the Chalmers Initiative of Cosmic Origins Postdoctoral Fellowship.
We thank the anonymous referee whose valuable comments helped improve the paper.
\end{acknowledgments}

\vspace{5mm}
\facilities{Atacama Large Millimeter/submillimeter Array (ALMA)}
\software{Common Astronomy Software Applications package \citep[CASA;][]{2022PASP..134k4501C}, CASSIS \citep{2015sf2a.conf..313V}}


\appendix

\section{Spatial distributions of each molecule} \label{sec:a1}

Figures \ref{fig:mom01}--\ref{fig:mom04} show moment 0 maps of molecular lines (contours) overlaid on continuum images (gray scale).
Red crosses indicate the positions of hot molecular cores (HMCs) identified in the moment 0 maps of CH$_{3}$CN.
If there is one HMC in a high-mass star-forming region, we did not indicate numbers.
We labeled numbers for HMCs in order of integrated intensity, from the highest to the lowest, if several HMCs are identified in a high-mass star-forming region.
Tables \ref{tab:mom0info} and \ref{tab:mom0info2} summarizes information on noise level and contour levels of each panel.
The same contour levels are applied for the same regions.
The line of H$_{2}$CO is contaminated with another line in two HMCs (G351.77-0.54 and NGC6334I), and we could not make its moment 0 maps.
 
\begin{figure*}[!th]
 \begin{center}
  \includegraphics[bb = 0 10 450 535, width=\textwidth]{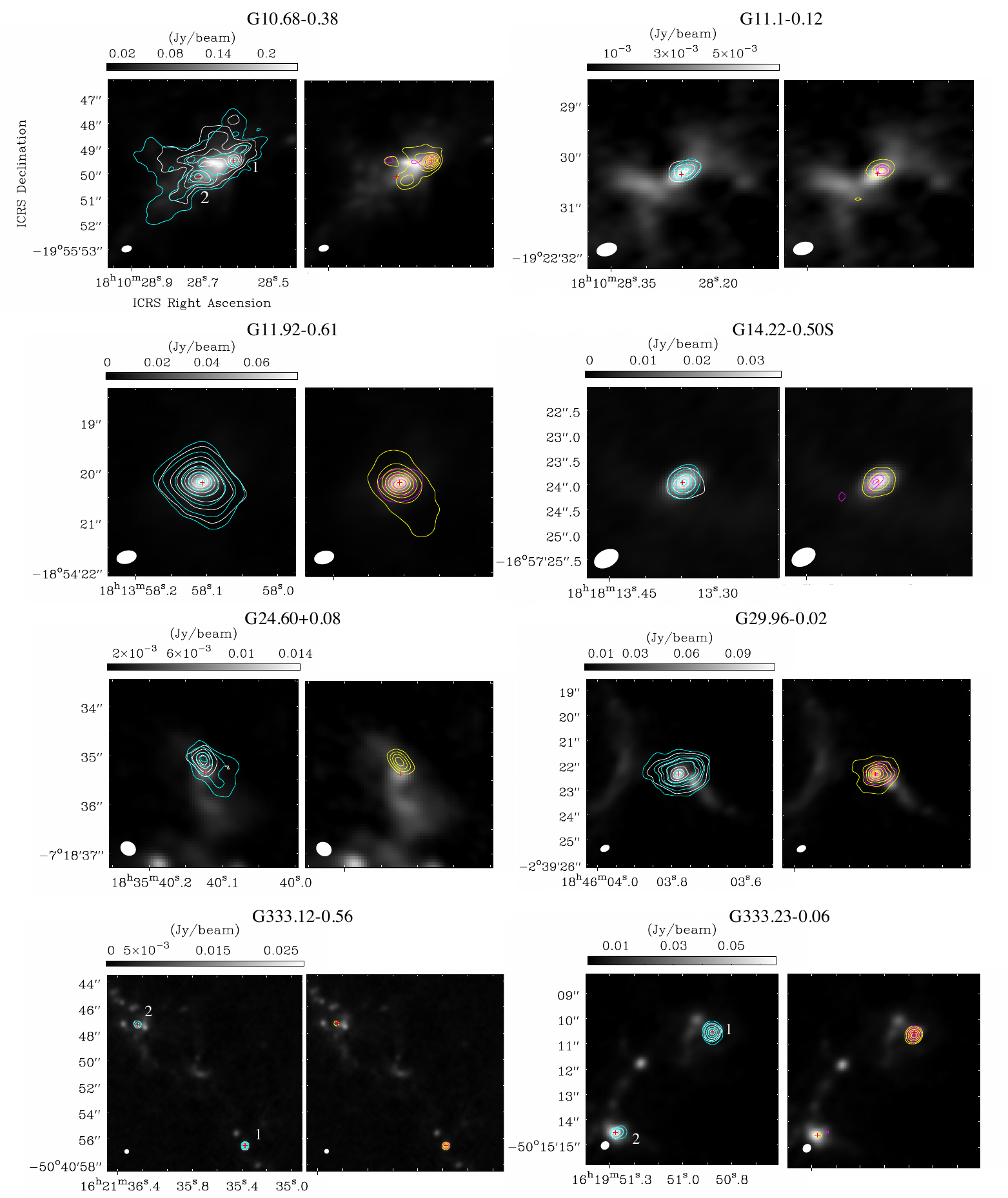}
 \end{center}
\caption{Continuum images (gray scales) overlaid with contours indicating moment 0 maps of molecular lines (left panels: white; CH$_{3}$CN and cyan; H$_{2}$CO, right panels: magenta; NH$_{2}$CHO and yellow; HNCO). Red crosses indicate positions of hot molecular cores (HMCs). Information on noise levels and contour levels are summarized in Tables \ref{tab:mom0info} and \ref{tab:mom0info2}. \label{fig:mom01}}
\end{figure*}

\begin{figure*}[!th]
 \begin{center}
  \includegraphics[bb = 0 10 465 542, width=\textwidth]{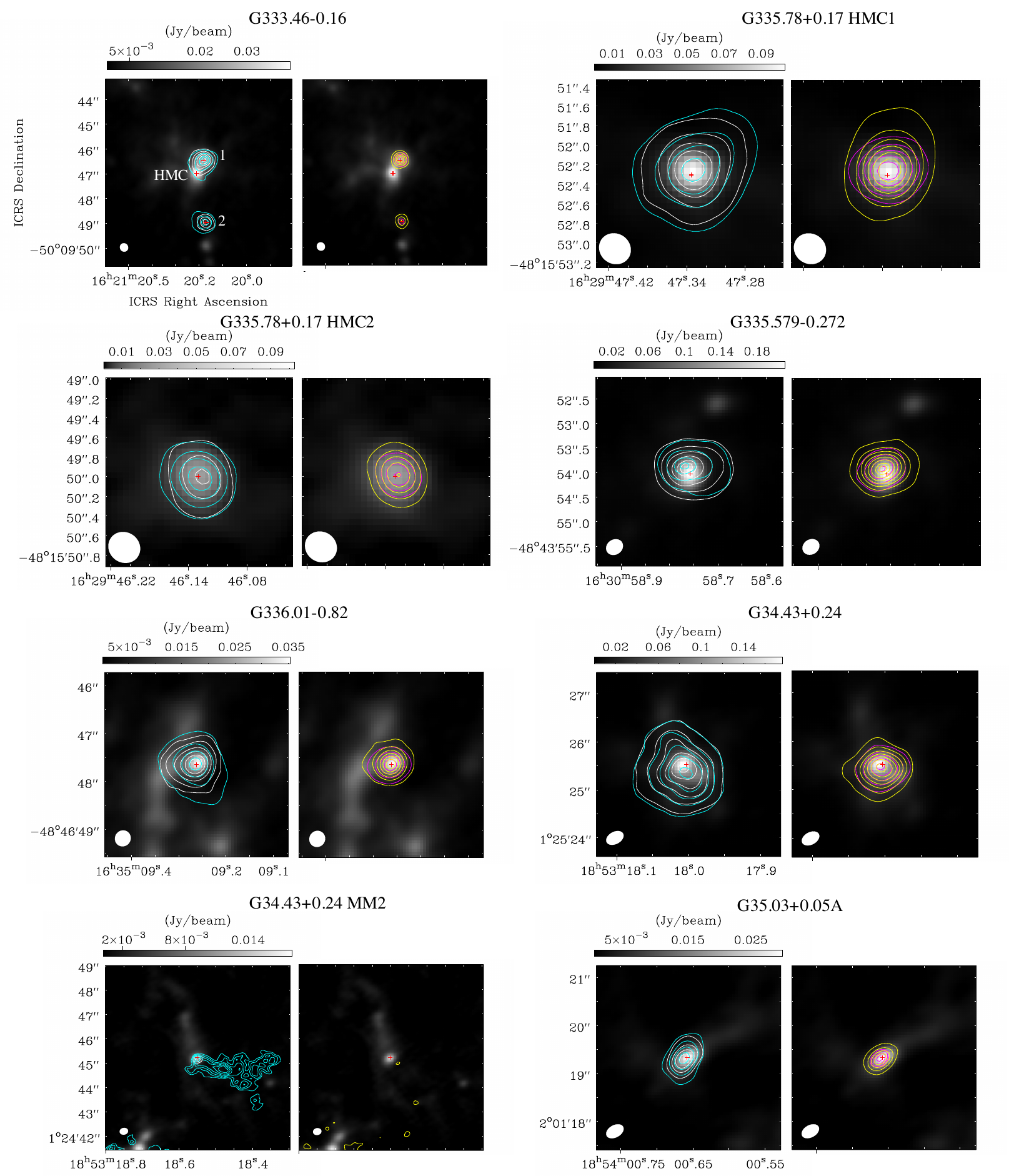}
 \end{center}
\caption{Continued.\label{fig:mom02}}
\end{figure*}

\begin{figure*}[!th]
 \begin{center}
  \includegraphics[bb= 0 10 453 541, width=\textwidth]{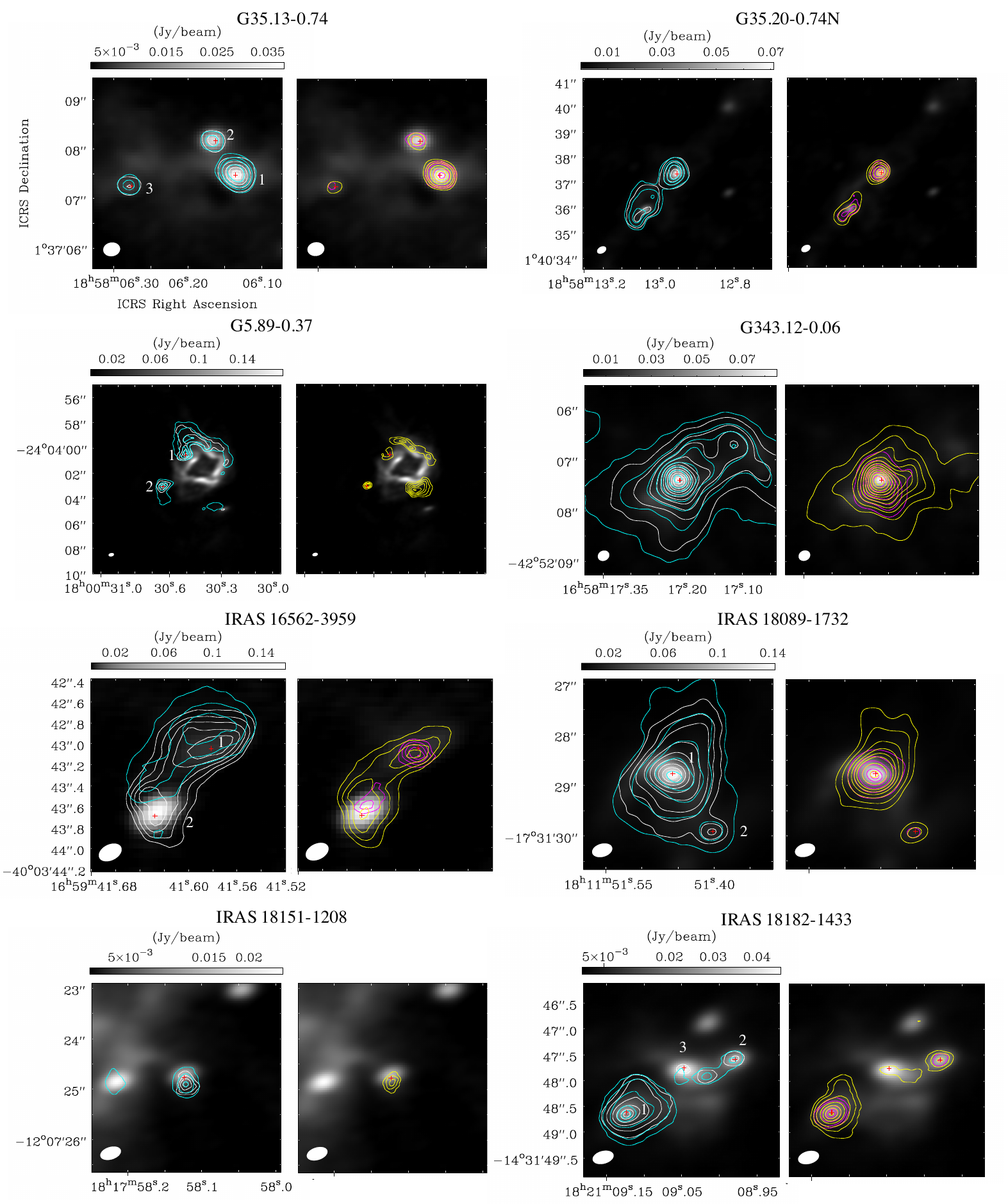}
 \end{center}
\caption{Continued.\label{fig:mom03}}
\end{figure*}

\begin{figure*}[!th]
 \begin{center}
  \includegraphics[bb= 0 10 462 419, width=\textwidth]{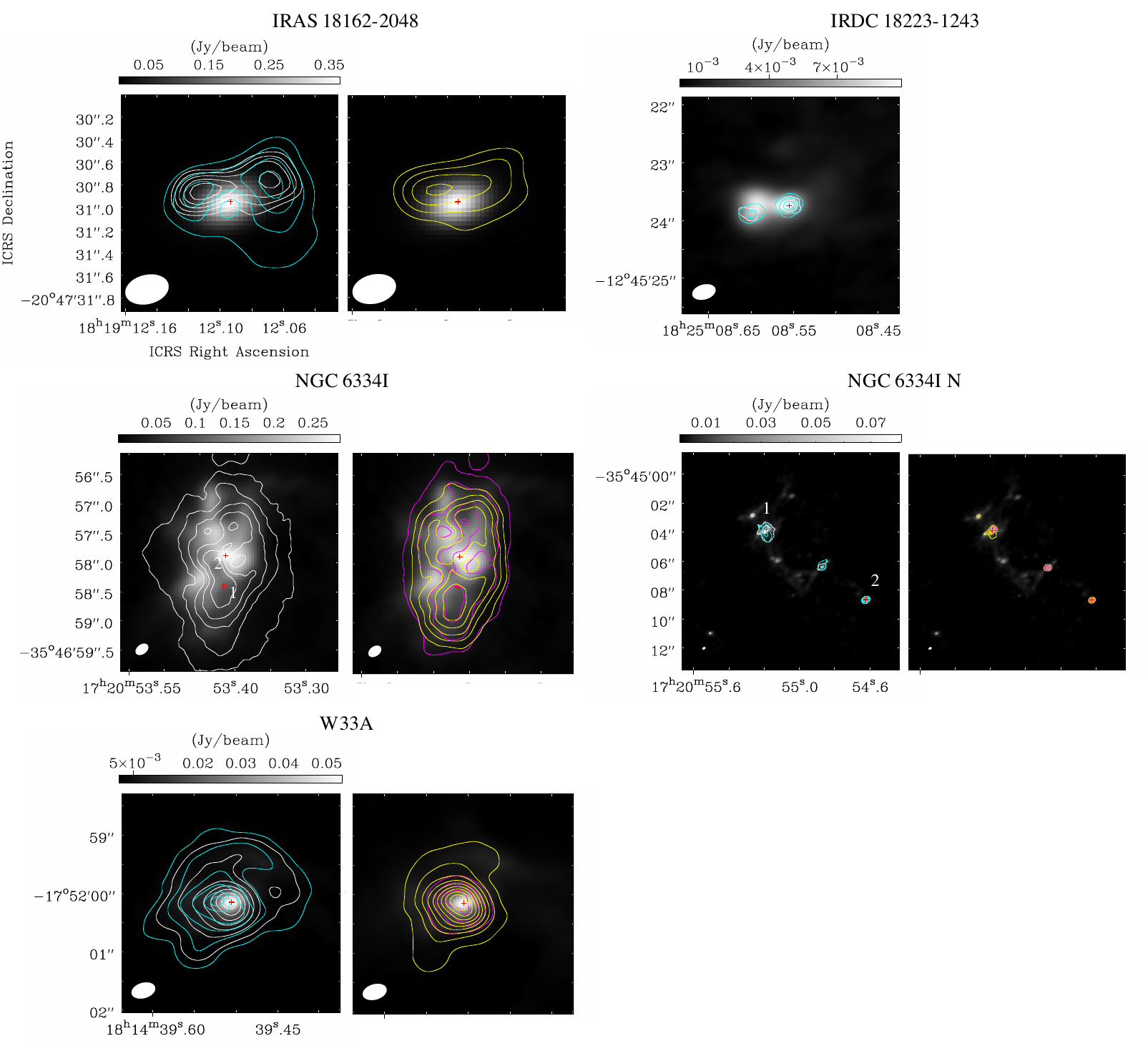}
 \end{center}
\caption{Continued.\label{fig:mom04}}
\end{figure*}

\begin{deluxetable*}{lllllllll}
\tabletypesize{\scriptsize}
\tablecaption{Information on noise level of continuum image and moment 0 maps \label{tab:mom0info}}
\tablewidth{0pt}
\tablehead{
\colhead{Region} & \colhead{Position} & \colhead{R.A.} & \colhead{Decl.} & \colhead{Continuum} & \colhead{CH$_{3}$CN} & \colhead{H$_{2}$CO} & \colhead{HNCO} & \colhead{NH$_{2}$CHO} 
}
\startdata
G10.62-0.38 & HMC1 & 18:10:28.61	 & -19:55:49.487 & 0.47 & 0.04	& 0.031	& 0.032	& 0.028	 \\
&HMC2 & 18:10:28.709 & -19:55:50.099	 \\
G11.1-0.12 &	HMC& 18:10:28.25	& -19:22:30.372	& 0.1 & 0.033& 0.025	& 0.025	& 0.026	 \\
G11.92-0.61	&HMC & 18:13:58.106 & -18:54:20.213 & 0.14& 0.043	& 0.031	& 0.034	& 0.028	 \\
G14.22-0.50 S & HMC & 18:18:13.348 & -16:57:23.955& 0.2 & 0.038 & 0.032 & 0.024 & 0.016 \\
G24.60+0.08 & HMC &18:35:40.124 & -7:18:35.356& 0.081 & 0.025 & 0.023 & 0.016 & ...  \\	
G29.96-0.02	& HMC &18:46:03.781 & -2:39:22.352 & 0.35 &0.072 & 0.041 &0.057 & 0.043  \\
G333.12-0.56 & HMC1 & 16:21:35.374 & -50:40:56.555 & 0.15 &0.036 & 0.032 & 0.028 & 0.045  \\
&HMC2 & 16:21:36.242 & -50:40:47.252 \\
G333.23-0.06	 & HMC1 & 16:19:50.875 & -50:15:10.526	&0.23 & 0.05 & 0.039	& 0.044 & 0.033  \\
&HMC2 & 16:19:51.276 &	-50:15:14.522 \\
G333.46-0.16 & HMC & 16:21:20.211 & -50:09:46.985 & 0.18 & 0.037	& 0.025	& 0.023	& 0.022	 \\
& HMC1 & 16:21:20.182 & -50:09:46.455 \\
& HMC2 & 16:21:20.171& -50:09:48.957	\\
G335.579-0.272& HMC & 16:30:58.758 & -48:43:54.011 & 0.45 & 0.052 & 0.036 & 0.039 & 0.038	  \\
G335.78+0.17 & HMC1 & 16:29:47.335 & -48:15:52.296 & 0.28 & 0.042 & 0.029 & 0.03 & 0.032 \\
&HMC2  &16:29:46.130 & -48:15:49.982 & 			&          &           &        &         \\
G336.01-0.82	 & HMC & 16:35:09.262& -48:46:47.638 & 0.21 & 0.043 & 0.024 & 0.034 & 0.029  \\
G34.43+0.24	& HMC & 18:53:18.005 & 1:25:25.524 & 0.39 & 0.067 & 0.045 & 0.047 & 0.04  \\
G34.43+0.24 MM2 & HMC & 18:53:18.552 &1:24:45.228 & 0.2	 & 0.019	& 0.024	& 0.012	 &  ... \\	
G35.03+0.35A & HMC &18:54:00.658& 2:01:19.330 & 0.16 & 0.056 & 0.033 & 0.038 & 0.031 \\
G35.13-0.74	& HMC1 & 18:58:06.135 & 1:37:07.476 & 0.17 & 0.03	& 0.03 & 0.024 & 0.02	  \\
&HMC2	&18:58:06.163&1:37:08.167\\
&HMC3	&18:58:06.278&1:37:07.247 \\
G35.20-0.74N & HMC & 18:58:12.952 & 1:40:37.357 & 0.22 & 0.043 & 0.037 & 0.043 & 0.036  \\
G351.77-0.54 & HMC & 17:26:42.531 & -36:09:17.376 & 1.0 & 0.059 & ...\tablenotemark{a} & 0.04 & 0.024\\
G5.89-0.37 & HMC1 & 18:00:30.507 & -24:04:00.561 & 0.44 & 0.058 & 0.061 & 0.042 & ... \\
& HMC2	& 18:00:30.639 & -24:04:03.082 \\
G343.12-0.06 & HMC & 16:58:17.212 & -42:52:07.402 & 0.19 & 0.036 & 0.024	& 0.025 & 0.019	 \\
IRAS 16562-3959	& HMC1 & 16:59:41.581 & -40:03:43.047 & 0.2 & 0.028 & 0.017 & 0.017 & 0.013 \\
& HMC2	& 16:59:41.627 & -40:03:43.691 \\
IRAS 18089-1732	& HMC1 & 18:11:51.457 & -17:31:28.771 & 0.23 & 0.044 & 0.025 & 0.029 &  0.026  \\
& HMC2 & 18:11:51.403 &	-17:31:29.919&&&&&\\
IRAS 18151-1208	 & HMC & 18:17:58.123 & -12:07:24.775 & 0.1 & 0.018 & 0.013 & 0.0087 & ...  \\	
IRAS 18182-1433	 & HMC1 & 18:21:09.123 & -14:31:48.644 & 0.14 & 0.025 & 0.019 & 0.022 & 0.015  \\
&HMC2	&18:21:08.979 & -14:31:47.590	 \\
&HMC3	&18:21:09.047 &-14:31:47.775 \\
IRAS 18162-2048	 & HMC & 18:19:12.093 & -20:47:30.946 & 0.32 & 0.033 & 0.033 & 0.027 & ... \\											
IRDC 18223-1243 & HMC &18:25:08.554 &-12:45:23.748 & 0.074	& 0.02 & 0.013 & ... & ...  \\		
NGC6334I & HMC1 & 17:20:53.416 & -35:46:58.397 & 1.2 & 0.072	& ...\tablenotemark{a} & 0.11 & 0.031  \\
& HMC2 & 17:20:53.413 & -35:46:57.881  \\
NGC6334I N	& HMC1 & 17:20:55.186 & -35:45:03.781 & 0.51 & 0.035 & 0.021 & 0.028 & 0.019  \\
& HMC2	&17:20:54.623 & -35:45:08.653	 \\
W33A & HMC & 18:14:39.505 & -17:52:00.147 & 0.17 & 0.047 & 0.029 & 0.033 & 0.026 \\
\enddata
\tablenotetext{a}{Moment 0 maps could not be made due to line contamination.}
\tablecomments{Units are mJy beam$^{-1}$ and Jy\,beam$^{-1}$ km\,s$^{-1}$ for continuum images and moment 0 maps of molecular lines, respectively. The position names and their coordinates are provided as a machine readable table.}
\end{deluxetable*}

\begin{deluxetable*}{llllll}
\tabletypesize{\scriptsize}
\tablecaption{Information on contour levels of continuum image and moment 0 maps \label{tab:mom0info2}}
\tablewidth{0pt}
\tablehead{
\colhead{Region} &  \colhead{CH$_{3}$CN} & \colhead{H$_{2}$CO} & \colhead{HNCO} & \colhead{NH$_{2}$CHO}
}
\startdata
G10.62-0.38 &   10-60 (10 step) &10,20,30 & 10-70 (10 step) & 10,20,30 \\
G11.1-0.12 &	5,10,15 & 10,15,20  & 5,10,15 & 4,5 \\
G11.92-0.61	&  10-150 (20 step) & 10,20,40,60,80,100 & 10-130 (20 step) & 10,30,50,70,80 \\
G14.22-0.50 S & 5,10,14 &	10,15,20,24 & 5,10,15 & 4,5 \\
G24.60+0.08 &  5,7,9,11 & 7,10,15,18 & 5,7,9,11,13 & ... \\	
G29.96-0.02	&10-130 (20 step) & 10-110 (20 step) & 10-110 (20 step) ,120 &10-90 (20 step) \\
G333.12-0.56 &10,20,30 & 10,15,20,25 &10,20,30 & 5,7,10 \\
G333.23-0.06	 &   10,15,25,35,45 &10,15,25,35,45 & 10,15,25,35,45 & 4-24 (4 step) \\
G333.46-0.16 &   10-60 (10 step) &	10,20,30,35 & 10-50 (10 step) & 10,20,30 \\
G335.579-0.272&   20-100 (20 step), 110	 & 20,40,60,70 &	10-130 (20 step), 140	& 10,30,50,70 \\
G335.78+0.17 HMC1 & 10-90 (20 step) & 10,30,50,70 & 10-110 (20 step) & 10-50 (10 step) \\
G335.78+0.17 HMC2	&	10,30,50 & 10,20,30,40 &	10,20,30,40,50 & 10,15,20,25 \\
G336.01-0.82	 &   10, 20-100 (20 step) & 10-70 (20 step) & 10-110 (20 step) & 10,30,50,70 \\
G34.43+0.24	&   10-110 (20 step) &	 10-90 (20 step) & 10-130 (20 step)	& 10-90 (20 step) \\
G34.43+0.24 MM2 &   5,6,7 &5,6,7,8,9,10 &	3,4,5 & ... \\	
G35.03+0.35A &   10,20,30,40 & 10,20,30,40 &10,20,30,40 & 10,12,15,18 \\
G35.13-0.74	&  10,20,30,40,50 &10,15,20,30 &10,15,20,30,40,50 & 5,7,10,20,30 \\
G35.20-0.74N &    10,30,50,70 & 10,20,30,40 & 10,20,30,50,70,80 & 5,10,20,30,40 \\
G351.77-0.54 &    10-230 (20 step) &	...\tablenotemark{a} & 10-210 (20 step)  & 10-90 (20 step) \\
G5.89-0.37 &   10,20,30,40 & 10,20,30,35 & 10,15,20,25,30,35 & ... \\
G343.12-0.06 & 10-230 (20 step) & 10-150 (20 step) & 10-230 (20 step) & 10-130 (20 step) \\
IRAS 16562-3959	& 10,15,20,25,30 & 10,15,20 & 10,15,20,25 & 5,7,9,11,13 \\
IRAS 18089-1732	& 10,20,30-150 (20 step) & 10,30,50,70,90,100 & 10, 20-100 (20 step), 150,200 & 10-90 (20 step), 100 \\
IRAS 18151-1208	 & 5,7,9,11 & 10,20,30 & 5,7,9 & ... \\	
IRAS 18182-1433	 &   10,20,40,60,80,90 &	10,15,30,50,60 & 5,10,20,40,60,80,100 & 5,10,30,50,60 \\
IRAS 18162-2048	 &10,12,14,16,18& 10,15,20,25 & 10,15,20,25 & ... \\											
IRDC 18223-1243 &5,7,9,11 & 10,15,20 & ... & ...  \\		
NGC6334I &    10,15,20,25,30& ...\tablenotemark{a} & 10,15,20,25,30& 	10,30,50,60 \\
NGC6334I N	&10,30,50 & 10,20,30 & 10,30,50 & 10,20,30 \\
W33A &10-130 (20 step),140 & 10-90 (20 step), 100 & 10-210 (20 step) & 10-110 (20 step) \\
\enddata
\tablenotetext{a}{Moment 0 maps could not be made due to line contamination.}
\end{deluxetable*}

\section{Velocity component and line width derived by the MCMC method} \label{sec:a2}

Table \ref{tab:vlsr} summarizes $V_{\rm {LSR}}$ and FWHM of each molecular line at each core derived from the MCMC method in the CASSIS software.
In the fitting, the initial guess of $V_{\rm {LSR}}$ is based on the CH$_{3}$CN line data, and we set the range of $V_{\rm {LSR}}=\pm 3$ km\,s$^{-1}$ from the initial guess in the MCMC analysis.

\begin{deluxetable*}{llccccccccccc}
\tabletypesize{\scriptsize}
\tablecaption{Velocity component and FWHM derived by the MCMC method \label{tab:vlsr}}
\tablewidth{0pt}
\tablehead{
\colhead{} & \colhead{} & \multicolumn{2}{c}{$^{13}$CH$_{3}$CN} & \colhead{} & \multicolumn{2}{c}{NH$_{2}$CHO} & \colhead{} & \multicolumn{2}{c}{HNCO} & \colhead{} & \multicolumn{2}{c}{H$_{2}$CO} \\
\cline{3-4} \cline{6-7} \cline{9-10} \cline{12-13}
\colhead{Region} & \colhead{Position} & \colhead{$V_{\rm {LSR}}$} & \colhead{FWHM} & \colhead{} & \colhead{$V_{\rm {LSR}}$} & \colhead{FWHM} &  \colhead{} & \colhead{$V_{\rm {LSR}}$} & \colhead{FWHM}  & \colhead{} & \colhead{$V_{\rm {LSR}}$} & \colhead{FWHM}
}
\startdata
G10.62-0.38 & HMC1 & 0.851 (3) & 5.041 (6) & & 1.71 (1) & 6.0 (4) & &  0.042 (1) & 5.866 (4) & & -1.318 (2) & 3.392 (4) \\
 & HMC2  & -3.194 (1) & 2.646 (3) & & ... & ... & & -2.805 (2) & 3.302 (4) & &  -3.142 (1) & 3.187 (3) \\
G11.1-0.12 & HMC & 31.85 (2) & 4.24 (1) & &  32.97 (7) & 3.4 (3) & &  31.404 (3) & 7.38 (5) &  & 31.57 (1) & 4.49 (3) \\
G11.92-0.61 & HMC &  35.1000 (3) & 10.8 (2) & & 35.6 (1) & 9.9 (8) & &  34.331 (1) &  13.594 (3) & & 34.8 (1) & 7.6 (5) \\
G14.22-0.50 S & HMC & 22.698 (8)\tablenotemark{a} &  6.23 (2)\tablenotemark{a}  & & 25.44 (3)  & 5.91 (2) & & 22.28 (1) & 9.07 (7) & & 23.070 (5) & 4.47 (1) \\
G24.60+0.08 & HMC & 52.5 (2)\tablenotemark{a} & 4.7 (4)\tablenotemark{a} & & ... & ... & & 54.05 (4) & 17.65 (8) & & 52.87 (1) &  7.19 (3) \\
G29.96-0.02 & HMC & 97.246 (1) & 7.976 (3) & & 97.097 (3) & 9.0 (1.2) & & 96.4 (5) & 9.0 (1.8)	& & 96.0 (2) & 8.29 (1) \\
G333.12-0.56 & HMC1 &  -58.47 (2) & 11.5 (3) & & -57.16 (2) & 9.77 (2) & &  -59.850 (5)	 & 11.52 (1) & & -60.56 (1)	 & 7.17 (3) \\
& HMC2	&  -54.10 (6) & 4.02 (1) & & -53.93 (1) & 6.03 (1) & & -55.167 (5) & 6.01 (1) & & -52.8 (2) & 4.9 (3) \\
G333.23-0.06 & HMC1 & -87.04 (5) & 4.3 (3) & & -86.5 (1) & 6.9 (3) & & -87.29 (4) & 7.3 (1.9) & & -87.49 (1) & 8.999 (1) \\
& HMC2 & -84.96 (2) & 4.1 (7)  & & -84.8 (3) & 5.5 (4) & &  -85.200 (1) & 9.96 (3) & & -85.12 (3) & 6.990 (8) \\
G333.46-0.16 & HMC & -44.057 (7) & 3.95 (2) & & ... & ... & &  -41.61 (1) & 8.9994 (3) & & -44.728 (4) & 3.54 (1) \\
& HMC1 & -43.55 (2) & 4.03 (1) & & -42.409 (3) & 5.271 (8) & & -43.261 (2) & 5.84 (5) & & -44.81 (1) & 2.0 (3) \\
& HMC2 & -39.23 (4)  & 3.9 (1)  & & -39.94 (4) & 7.4 (4)  & &  -42.446 (5) & 9.74 (1)  & & -40.997 (2) & 7.72 (2) \\
&&  -45.15 (2)\tablenotemark{b} & 3.87 (9)\tablenotemark{b} & & &   & &   &  & & &  \\
G335.579-0.272 & HMC & -46.44 (7) & 3.2 (7) & & -45.82 (4) & 5.5 (6) & &  -47.21 (1) & 6.7 (1.0) & & ... & ...  \\
G335.78+0.17 & HMC1 & -48.94 (6) & 8.304 (4) & & -48.990 (3) & 9.144 (9) & & -51.1 (1) & 10.0 (1.1) & & -49.37 (6) & 11.999 (4) \\
& HMC2 & -50.51 (9) & 7.8 (1) & & -50.769 (5) & 8.51 (1) & & -51.80 (7) & 8.4 (1.1) & & -50.654 (5) & 8.34 (1) \\
G336.01-0.82 & HMC & -47.186 (8) & 10.01 (9) & & -45.665 (2) & 10.492 (5) & & -46.2 (2) & 9.9 (2) & & ... & ... \\
G34.43+0.24 & HMC & 58.440 (2) & 7.5 (2) & & 59.3 (2) & 6.11 (7) & & 58.5 (2) & 8.8 (4) & & ... & ... \\		
G34.43+0.24 MM2 & HMC & 55.502 (2)\tablenotemark{a} & 4.8 (1)\tablenotemark{a} & & ... & ... & & 55.86 (4) & 3.61 (6) & & 55.41 (2) & 7.988 (9) \\
G35.03+0.35A & HMC & 45.94 (2) & 5.91 (7) & & 45.342 (2) & 4.14 (4)  & & 45.71 (1) & 6.4 (2.2) & & 47.733 (3) & 7.140 (6)  \\
G35.13-0.74 & HMC1 & 35.93 (2) & 7.67 (5) & & 31.57 (2) & 7.0 (2)  & & 31.6 (4) & 2.6 (9)  & & ... & ... \\	 
&&&& & 38.00(1)\tablenotemark{b} & 4.14 (5)\tablenotemark{b} & & 36.73 (4)\tablenotemark{b} & 3.6 (6)\tablenotemark{b} & & ... & ... \\
& HMC2	& 34.39 (2) & 8.54 (9)  & & 36.1 (1) & 7.1 (6) & & 34.031 (9) & 11.00 (3) & & 34.585 (9) & 7.85 (2) \\
& HMC3 & 35.527 (6)\tablenotemark{a} & 4.65 (4)\tablenotemark{a} & & 33.91 (3) & 8.00 (1) & & 34.758 (5) & 6.16 (2) & & 36.205 (5) & 5.23 (1) \\
G35.20-0.74 N & HMC & 32.113 (3) & 5.165 (8)  & & 32.13 (4) & 6.4 (5) & & 31.602 (3) & 6.973 (5) & & 31.43 (9) & 4.66 (9) \\
G351.77-0.54  & HMC & -3.825 (1) & 9.127 (3)  & &  -2.538 (1) & 6.708 (4)  & & -2.53 (8) & 2.9 (6)  & & ... & ... \\				
G5.89-0.37 &	HMC1 & 11.132 (3) & 5.24 (1)  & & ... & ... & & 11.808 (4) & 6.825 (8) & & 11.5 (2) & 7.5 (3) \\
& HMC2 & 7.918 (9) & 6.47 (2) & & ... & ... & & 8.54 (3) & 7.13 (3) & & 9.189 (2)  & 5.529 (5) \\
G343.12-0.06 & HMC & -33.708 (3) & 9.09 (4) & & -32.77 (3) & 5.02 (7)  & & -32.01 (1) & 6.2 (5) & & -33.124 (5) & 7.215 (6) \\
IRAS 16562-3959	& HMC1 & -14.233 (1) & 3.897 (6) & & -13.222 (1) & 3.536 (3) & &  -13.722 (4) & 3.176 (2) & & -14.90 (5) & 3.1 (3) \\
& HMC2	& -16.86 (2) & 4.9998 (2) & & -17.142 (4) & 6.482 (8)  & &  -17.3 (2) & 6.7 (2)  & & -17.207 (5) & 6.54 (1) \\
IRAS 18089-1732	& HMC1 & 32.875 (3) & 7.88 (8)  & & 32.36 (4) & 7.5 (3)  & & 31.55 (3) & 7.8 (1.8)  & & 34.20 (1) & 5.8 (3) \\
& HMC2 & 33.02 (6) & 8.0 (1) & & 33.399 (1) & 10.67 (5) & & 33.172 (6) & 10.88 (1) & & 33.743 (3) & 6.177 (7) \\
IRAS 18151-1208	 & HMC &  30.410 (2)\tablenotemark{a} & 2.95 (2)\tablenotemark{a} & & ... & ... & & 30.073 (5) & 3.48 (1) & & 30.520 (2) & 3.030(5) \\	
IRAS 18182-1433	 & HMC1 &  61.12 (2) & 8.51 (3) & & 61.7 (3) & 6.9 (1.8) & & 59.8 (2) & 9.9 (2) & & 60.034 (2) & 7.647 (5) \\
&HMC2	& 60.635 (8) & 4.67 (1) & & 61.0 (1)  & 6.1 (1.2) & & 60.59 (3) & 5.3 (1.1) & & 61.038 (4) & 5.20 (1) \\
&HMC3	& 62.78 (1) & 4.64 (3)  & & ... & ... & & 61.94 (2) & 7.9994 (5) & &  63.398 (8) & 2.9998 (2) \\
IRAS 18162-2048	 & HMC & 13.816 (2)\tablenotemark{a} & 5.843 (6)\tablenotemark{a} & & ... & ... &  & 13.8 (3) & 4.5 (1.0) & & 14.745 (2) & 2.561 (6) \\										
IRDC 18223-1243 & HMC & 45.685 (8)\tablenotemark{a} & 6.55 (5)\tablenotemark{a} & & ... & ... &  & 45.84 (2) & 9.05 (6) & & 45.532 (5) & 5.35 (2) \\		
NGC6334I N	& HMC1 & -1.768 (2) & 7.37 (1) & & -0.699 (3) & 7.44 (1) & & -2.6 (7) & 7.0 (1.6)  & & ... & ... \\
& HMC2	& -7.59 (6) & 3.6 (2)  & & -5.547 (3) & 10.529 (6) & & -6.47 (6) & 6.6 (1.2) & & -7.47 (2) & 4.374 (7) \\
W33A & HMC & 37.939 (3) & 8.55 (2) & & 39.53 (2) & 9.32 (1) & & 38.51 (2) &  6.3 (5) & &  38.802 (1) & 6.938 (3)  \\
\enddata
\tablecomments{Unit is km\,s$^{-1}$. The numbers in parentheses indicate the standard deviation derived from the MCMC analysis, expressed in units of the last significant digits.}
\tablenotetext{a}{Derived from fitting of the lines of CH$_{3}$CN.}
\tablenotetext{b}{Applied by two velocity-component fitting.}
\end{deluxetable*}

\section{Comparison with peak abundances of Model B} \label{sec:a3}

\begin{figure*}[!th]
 \begin{center}
  \includegraphics[bb = 0 20 545 430, scale = 1.0]{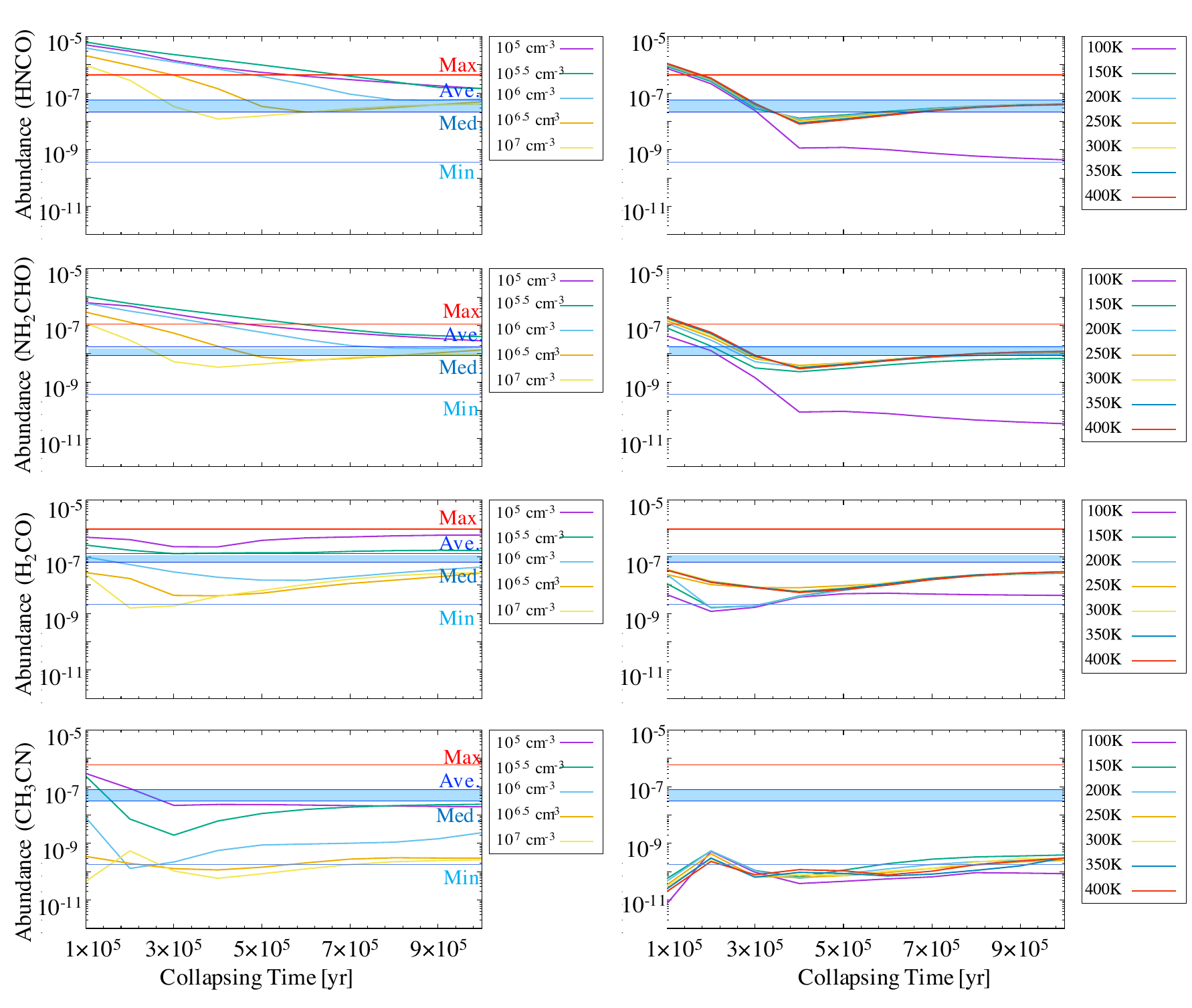}
 \end{center}
\caption{Comparison with maximum abundances of Model B \citep{2020ApJ...895...86G}. Panels from top to bottom show results of HNCO, NH$_{2}$CHO, H$_{2}$CO, and CH$_{3}$CN, respectively. Left panels show dependences on different collapsing timescale and maximum density. Right panels show dependences on different collapsing timescale and maximum temperature with the maximum density of $10^7$ cm$^{-3}$. The four representative observed abundances are plotted (Maximum, Average, Median, and Minimum). The blue filled ranges indicate the ranges between average and median values. \label{fig:modelBpeak}}
\end{figure*}

Figure \ref{fig:modelBpeak} shows comparisons of observational results with the modeled maximum abundances obtained by Model B \citep{2020ApJ...895...86G}.

\bibliographystyle{aasjournal}
{}



\end{document}